\documentclass[12pt]{article}
\usepackage{makeidx}
\usepackage{epsfig}
\usepackage{epstopdf}
\usepackage{amsmath}
\usepackage{latexsym}
\usepackage{textcomp}
\usepackage{amssymb}
\usepackage{amsfonts}
\usepackage{amsmath}
\usepackage[all]{xy}
\usepackage{amsfonts}
\usepackage{amssymb,amsmath}
\usepackage{color}
\usepackage{caption}
\usepackage{subcaption}
\usepackage{comment}

\usepackage[margin=0.7in]{geometry}
\usepackage[singlespacing]{setspace}

\newtheorem{defin}{ {Definition}}[section]
\newtheorem{theorem}[defin]{\bf {THEOREM}}
\newtheorem{lemma}[defin]{\bf {Lemma}}

\newtheorem{prop}[defin]{\sc {Proposition}}

\newcommand{\proof}{\noindent{\bf Proof: }}

\newcommand{\QED}{\hfill \mbox{\rule{2mm}{2mm}} \vspace{0.5cm}}

\newcommand{\Xomit}[1]{}

\makeindex
\begin{document}
\title{The Value of Sharing Intermittent Spectrum\thanks{This work
was supported by NSF under grant AST-134338.}}
\author{R. Berry\footnote{EECS Department, Northwestern University,
Evanston, IL 60208}, M. Honig$^\dagger$, T. Nguyen\footnote{Krannert School of Management, Purdue University}, V. Subramanian\footnote{EECS Department, University of Michigan} \& R. V. Vohra\footnote{Department of Economics and Department of Electrical and Systems Engineering, University of Pennsylvania.}}
\date {}
\maketitle

\abstract{Recent initiatives by regulatory agencies to increase spectrum resources
available for broadband access include rules for sharing spectrum with high-priority incumbents.
We study a model in which wireless Service Providers (SPs) charge for access to their
own exclusive-use (licensed) band along with access to an additional shared band.
The total, or delivered price in each band is the announced price plus a congestion cost,
which depends on the load, or total users normalized by the bandwidth.
The shared band is intermittently available with some probability, due to incumbent activity,
and when unavailable, any traffic carried on that band must be shifted 
to licensed bands. The SPs then compete for quantity of users.
We show that the value of the shared band depends on the relative sizes of the SPs:
large SPs with more bandwidth are better able to absorb the variability caused
by intermittency than smaller SPs. However, as the amount
of shared spectrum increases, the large SPs may not make
use of it. In that scenario shared spectrum creates more value
than splitting it among the SPs for exclusive use. We also show
that fixing the average amount of available shared bandwidth,
increasing the reliability of the band is preferable to increasing the bandwidth.}
%fixing the average amount of shared band $\alpha W$,
%the value of the shared band generally increases by reducing the bandwidth $W$
%and increasing the reliability $\alpha$.}

\section{Introduction}
The evolution of wireless networks for mobile broadband access has led
to a proliferation of applications and services that have greatly increased the demand
for spectrum resources. In response, regulatory agencies have introduced 
new initiatives for increasing the amount of spectrum that can be used to meet this demand.
These include auctions for repurposing bands previously designated for restricted use,
such as broadcast television, and also initiatives for sharing spectrum
assigned to government agencies. 
Proposed methods for sharing spectrum were highlighted in the 2012 report by
the President's Council of Advisors on Science and Technology (PCAST) \cite{PCAST}.
Sharing is motivated by the recognition that relocating the associated
services (e.g., satellite) to other bands would be expensive 
and incur large delays, and that much of the federal spectrum is used sporadically
and often only in isolated geographic regions.

Our objective in this paper is to provide insight into the potential benefits of sharing
spectrum that is intermittently available. We take into account the congestion
caused by sharing along with strategic decisions made by competing Service Providers (SPs).
Here we are not concerned with the incentives needed for the incumbent 
Federal agencies to make more efficient use of their spectrum, which could
include sharing, but rather assume that a given amount of spectrum 
is available for sharing.\footnote{For example, an incentive for providing shared access,
proposed in the PCAST report,
might be a `scrip' system that rewards more efficient use of spectrum.
Another approach, discussed in \cite{skorup2015},
is to assign overlay rights as an alternative to sharing.}
We then compare the value of the band when shared among all SPs
versus a partition and assignment of the sub-bands among the SPs 
for their exclusive use.\footnote{``Exclusive use''
spectrum is also shared, of course, but only by customers of the assigned SP.}

%Much of the spectrum currently allocated to federal agencies 
%has the following properties: Relocating the associated
%services (e.g., satellite) to other bands would be expensive 
%and incur large delays; the spectrum is used sporadically
%and often only in isolated geographic regions.
%This motivated the recommendation in the
%2012 Presidential Council of Advisors on Science and Technology 
%(PCAST) report \cite{PCAST} that
%up to 1000 MHz of federal spectrum be {\em shared} with commercial users. 
%The FCC has been developing rules to promote sharing 
%including the 2010 rules for the TV white spaces \cite{FCC10174}
%as well as the proposed rules for sharing spectrum
%in the 1.7 GHz band, and ``light" licenses for deploying
%small cells in the 3.5 GHz band \cite{FCC-12-148}.

%We take as given the recommendation made in the 2012 PCAST report that Federal spectrum would be made available under a simple priority rule:
We present a model in which multiple wireless SPs compete 
to provide service to a pool of nonatomic users. Each SP has its own {\em proprietary} band
for exclusive use, and in addition, there is a band assigned to another incumbent
(e.g., Federal agency) that can be shared. The incumbent has priority in the shared band
and can pre-empt secondary users. We model this by assuming that
the shared band is available with probability $\alpha$. We compare two modes
for sharing: {\em licensed} sharing in which the shared bandwidth is partitioned
among the SPs as additional proprietary, but intermittent bandwidth, 
and {\em authorized open access}
in which the set of designated SPs all share the entire band with the incumbent.
In both cases the SPs compete for users according to a Cournot game.
Specifically, each SP chooses a quantity of users, which it can allocate
across its proprietary and shared bands. The total, or delivered price in each band
paid by a user, determined by a linear demand function,
is an announced price plus a congestion, or latency cost, which increases linearly with load,
or users per unit bandwidth. We compare the equilibrium social welfare 
and consumer surplus for both licensed and open access sharing.

Our model can be interpreted in the context of the sharing rules 
recently approved by the US Federal Communications Commission 
for the Citizens Broadband Radio Services (CBRS) band 
from 3550 to 3650 MHz (100 MHz total).\footnote{The incumbent service in this band is Naval radar,
which is primarily active along the east and west coasts of the US.
The band is currently used sporadically with some predictability
about availability. To coordinate use of the band between Naval radar
and secondary lower-priority users, the secondary users must register with a database that
updates and authorizes activities within the band.}
The rules allow for licensed shared access, corresponding to our licensed model
for the shared band, as well as ``general authorized access'', corresponding
to our open access model. 
Our model allows for splitting the shared band between these two modes. 
We emphasize, however, that ``open access"
is restricted to the given set of competing SPs.
We do not consider the possibility of open access competition from additional SPs
that do not have access to their own exclusive use (proprietary) bands.
For both access modes, the spectrum is provided 
at a granularity of 10 MHz channels.
Therefore, when the spectrum is unavailable, the entire band is unavailable.
In that scenario we assume that all traffic assigned to the shared band
by the SPs must be diverted to their respective proprietary bands.

Comparing the equilibria associated with licensed and open access sharing,
we find that which method should be used depends on the structure of the market. 
Intuitively, open access bandwidth should lead to greater congestion and this is indeed the case. However, if there is sufficient competition, there will also be lower prices. Will prices fall sufficiently to offset the effect of greater congestion? We find that if the market consists of a large number of SPs, no one being dominant in terms of the amount of their proprietary bandwidth, the open access regime generates more consumer surplus than the licensed regime. Total welfare (consumer surplus
plus SP revenue), however, is higher in the licensed regime than for open access.
This difference arises from lower prices in the open access regime but higher latency. 

If there is a limit to how much congestion consumers will tolerate, one might consider 
a `mixed' policy where the shared bandwidth is split between licensed and open access. 
As the fraction of licensed bandwidth increases, the SPs shift traffic from the open access band
to their licensed bands, raising prices for the licensed bands.
As licensed sharing is initially introduced, relative to full open access for the entire shared band,
the {\em average} price (across proprietary and shared bands)
{\em and} average latency initially {\em increase}.
However, once a critical threshold on the amount of licensed bandwidth is exceeded, 
prices continue to increase but latency drops. The observed net effect
is that consumer surplus decreases while social welfare increases.

The licensed regime generates greater social welfare provided there is sufficient competition. 
What if this is not the case? We model this possibility as a market that contains one or more SPs characterized by a relatively large amount of proprietary bandwidth. This leads to a tradeoff:
larger SPs are better able to handle the intermittency associated with the shared band,
but allocating more licensed bandwidth to the larger SPs places the smaller SPs
at a disadvantage, compromising the benefits of competition. When there are only a few
large SPs, and low intermittency, consumer surplus therefore benefits the most from
allocating more shared bandwidth to the smaller SPs. Interestingly,
if the large SP has sufficient proprietary bandwidth, allocating the shared band
as open access achieves the same outcome, namely, the large SP
does not make use of the open access spectrum due to the congestion from the other SPs.
%That in turn alleviates congestion in the open access spectrum increasing its value.
%We find that the
%open access regime generates more consumer surplus than  regime. This is because a large SP is better able to absorb the variability associated with intermittently available bandwidth. Therefore, in the open access regime, highly intermittent additional bandwidth will place the smaller SPs at a disadvantage relative to the larger SPs. However, if a large SP has 

If the shared bandwidth is to be licensed, how should it be allocated among the SPs? Auctions are the standard response. We find that the natural auction rule for allocating licensed bandwidth, giving it all to the highest bidder, will be inefficient. The problem is that the marginal benefit of additional bandwidth  increases with the amount of proprietary bandwidth that each bidder possesses prior to auction. This comes from the fact that such a bidder is better able to absorb the variability associated with intermittently available bandwidth. Thus, bidders endowed with a larger amount of initial bandwidth are willing to pay more for additional bandwidth. This produces a lop sided distribution of bandwidth which reduces consumer {\em and} social welfare. However, if the shared band is open access, 
a large SP may not use it, leaving it for smaller SPs. In fact, we give an example where 
given a choice between open access and licensed access allocated by auction, perhaps surprisingly,
the bidders would strictly prefer open access.

We also consider the tradeoff between the reliability of the shared band and
the amount of shared bandwidth. Holding the expected quantity of shared bandwidth
fixed (i.e., $\alpha W$ where $W$ is the shared bandwidth),  
SPs would prefer a smaller amount of shared bandwidth with greater availability.
The variation in how the shared band is valued can vary substantially with $\alpha$,
depending on the relative amounts of proprietary bandwidth.

\subsection{Related work}
This paper fits within the stream of work that analyzes the impact of spectrum policy using models of competition with congestion costs. Examples of such models can be found in \cite{AO07}, \cite{AF07} and \cite{JohariWR10}. It differs from prior work in this stream with its focus on intermittently available spectrum. A similar Cournot competition model with congestion has been studied in \cite{perakis2014efficiency}; here we enrich that model by allowing an additional shared resource 
along with intermittency.

The paper closest to this one is \cite{nguyen2014cost}.
There the shared band is {\em non-intermittent}, and the SPs compete
according to a Bertrand model. That model is motivated by the scenario
in which the shared band can be designated as unrestricted open access.
Price competition, as opposed to quantity competition, better fits
the scenario in which the traffic assigned to the open access band
always stays in that band. There the equilibrium price in the shared open access band
is shown to be zero. Although not explicitly modeled in \cite{nguyen2014cost},
that also reflects the scenario in which there may be
additional competition from entrants with no proprietary spectrum.
In contrast, for the Cournot model considered here, the price
of the open access band is typically strictly positive reflecting
the potential cost of having to carry the traffic in proprietary bands.
Another difference is that here larger SPs are at an advantage 
because they are better able to handle intermittency.

The remainder of this paper is organized as follows.
The next section presents the Cournot model with congestion. 
In Section~\ref{sec:TwoSPs} we analyze this model for the case of two competing SPs.
We think of this as modeling the scenario with an oligopoly of wide-area cellular SPs.
Section \ref{sec:manySPs} examines the opposite case with many SPs,
each with a proportionately  
small share of the available proprietary bandwidth.
This models the scenario in which there are low barriers to entry
so that many operators may wish to set up competing networks within a local area.
Numerical results are presented that illustrate the tradeoffs among 
competition, prices, and latency.
In all cases we compare consumer surplus and social welfare under
different assumptions concerning the amount of available licensed versus 
available shared bandwidth.
Section \ref{sec:marginal} presents extensions to concave decreasing demand 
and convex increasing latencies. 
Section \ref{conc} concludes and proofs of the main results are given in the appendices.

%In Section \ref{InfSPs} we then consider the opposite extreme in which there are a large number of providers, again with  different assumptions concerning amounts of proprietary and shared spectrum. 

\section{The Model}\label{model}
Suppose $N$ SPs  compete to offer wireless service to a common pool of customers.  Each SP $i \in \{1,2,\ldots N\}$ possesses an amount of proprietary (licensed) bandwidth, denoted $B_i$. In the status quo, this is the only resource the SPs can access. We are interested in the scenario
where an amount $W$ of new spectrum is made available that is to be shared
with an incumbent user. 
%The primary user of this band is a government agency. 
When the incumbent is actively using the band, it is unavailable to carry traffic for the SPs. 
Otherwise, it is available for the SPs.
We consider two different policies that govern the way a particular SP can access this band: 
{\it licensed access}, where a part of the band $W_i$ is designated for exclusive use
by SP $i$, and
%the secondary band is licensed to a single SP 
%for secondary use and 
{\it open access} where all SPs can access the band. 
We will allow the shared band to be divided 
into several disjoint sub-bands, where each sub-band can
be designated as either licensed to a particular SP, or as open access.
To simplify the model description, we first assume that the entire shared band is either
licensed to a single SP, or is open access. We subsequently
consider the scenario in which parts of the shared band are allocated
to different SPs for licensed and open access.

%{\bf Also say something about in licensed case we allow the band to be divided up among %multiple SPs. In 3.5 GHz - the plan is to do a combination of these - part of the band is %unlicensed and part is licensed - is there something here to be looked at?}

We assume a pool of infinitesimal customers or users with a downward slopping inverse demand curve 
\begin{equation}
P(y) = 1-y,
\label{Py}
\end{equation}
which gives the marginal utility obtained by the $y$th customer served, 
where all customers require the same amount of (average) service. 
As in \cite{nguyen2014cost}, the price the $y$th customer is willing 
to pay for service is given by the difference between their marginal utility 
and the latency or congestion cost they experience. 
Following \cite{nguyen2014cost}, we refer to the sum of the latency cost 
and the service price as the {\it delivered price}. 

If SP $i$ serves $x_i$ customers on its  proprietary band, the resulting latency cost is given by 
\begin{equation}
\ell_i(x_i)=\frac{x_i}{B_i},
\label{elli}
\end{equation}
which is increasing in the amount of traffic served 
and decreasing in the amount of bandwidth available to the SP. 
When each SP $i$ serves $w_i$ customers using the entire band of secondary spectrum, 
we model the latency by 
\begin{equation}
\ell_w\left(\sum_i w_i\right) = \frac{1}{W} \sum_i w_i,
\label{ellw}
\end{equation}
which is now increasing in the sum of the traffic from the SPs 
and decreasing in the available secondary spectrum $W$. 
Note, if the entire secondary band is licensed to a single SP $i$,
this corresponds to constraining $w_j=0$, $j\neq i$. 
We assume that the shared band is intermittently available 
%To model the possibility that this secondary spectrum is intermittently available, assume it is available 
with probability $\alpha \in [0,1]$.
When unavailable, the traffic designated by SP $i$ for the secondary band, 
$w_i$, must be off-loaded onto SP $i$'s proprietary band.\footnote{This is a 
reasonable assumption when the customers have a high dis-utility 
for not receiving service.} Thus, the `expected' latency of traffic 
served by SP $i$ on its proprietary band is
\begin{equation}
\bar{\ell}_i = (1-\alpha)\cdot \ell_i(x_i+ w_i)+ \alpha \cdot \ell_i(x_i).
\label{barelli}
\end{equation}
The `expected' latency of traffic  experienced by SP $i$'s traffic on the secondary spectrum will be
\begin{equation}
\bar{\ell}_w =
\alpha\cdot \ell_w\left(\sum_i w_i\right)+ (1-\alpha)\cdot \ell_i(x_i+w_i).
\label{barellw}
\end{equation}
Proprietary spectrum is assumed to be available at all times.\footnote{As we show below, our model can also be applied to the case where some portion of the  proprietary spectrum 
is intermittently available, provided that the remainder is always available.}

The SPs compete according to a Cournot model.\footnote{FCC report 10-81 \cite{FCC10-81} 
states that mobile wireless SPs compete on dimensions other than price.
In particular, a network upgrade can be interpreted as an attempt to expand capacity.}
Each SP $i$ decides on a pair $(x_i,w_i)$, which represents the amount of traffic it will carry. Given a choice of $(x_i, w_i)$ by each SP $i$, the resulting price paid by the users will be the difference between their marginal utility and the resulting expected latency. Specifically, the delivered price for the user load is
\begin{equation}
p_d = P\left(\sum_i (x_i + w_i)\right)
= 1- \sum_i (x_i + w_i).
\label{pd}
\end{equation}
The actual price paid for service depends on the latency experienced
by the traffic. For SP $i$'s licensed band, the price is given by
\begin{equation}
p_i= p_d -(1-\alpha)\cdot \ell_i(x_i+w_i)- \alpha \cdot \ell_i(x_i),
\label{pi}
\end{equation}
and for the secondary band, the price paid by SP $i$'s users is given by
\begin{equation}
p^w_i= p_d -\alpha\cdot \ell_w\left(\sum_i w_i\right)-(1-\alpha)\cdot \ell_i(x_i+ w_i).
\label{pwi}
\end{equation}
Each SP $i$ seeks to maximize its revenue given by
\begin{equation}
R_i = p_ix_i + p^w_i w_i.
\label{pii}
\end{equation}

The model just described assumes that each SP serves two classes of customers: one using their proprietary band and the other with the shared spectrum, charging each class different prices. However, one can also interpret the model as one where there is only one class of customers and the SP decides whether to serve each customer via the proprietary or secondary bands. Formally, we think of $\frac{x_i}{x_i+w_i}$ and $\frac{w_i}{x_i+w_i}$ as the probability that a consumer is served via proprietary spectrum or secondary spectrum, respectively. The price that SP $i$ charges is then
\begin{equation}
\bar{p}_i = \frac{x_i}{x_i+w_i}p_i+ \frac{w_i}{x_i+w_i}p_i^w.
\end{equation}
The revenue of SP $i$ is still given by \eqref{pii}.

\subsection{Shared Sub-bands}

In the preceding model the entire band of shared spectrum is either 
licensed to one SP, or is open access. 
More generally, we allow this band to be divided into multiple disjoint 
sub-bands with bandwidths $W_k\geq 0$, $k=0,1,\ldots, N$, with
\[
\sum_{i=0}^N W_k = W.
\]
Here, $W_i$ represents the part of the shared band allocated to user $i$
as licensed bandwidth and $W_0$ represents any remaining bandwidth
that is allocated for open access
(where any of these terms may be zero if no bandwidth is allocated in that way).  
The resulting traffic load for a sub-band with $W_i$ units of bandwidth is then given by 
$
%\ell_{W_i}(y) = 
%\frac{y}{W_i}
y/W_i
$
where $y$ is the total traffic in that sub-band.
We assume that when the incumbent is active, it claims the {\em entire} band
so that all sub-bands must be vacated.

Following the preceding model, a SP would then specify an amount of traffic 
for each shared sub-band it is permitted to use, as well as for its proprietary band.
However, the next result shows that we can `pool' all of the licensed bands 
assigned to an SP and represent them as a single equivalent band 
that serves the aggregate traffic on these sub-bands. 
Formally, an SP with proprietary bandwidth $B_i$ and 
licensed shared bandwidth $W_i$  can be viewed as having a single band
having bandwidth $B_i + W_i$ with probability $\alpha$ and 
bandwidth $B_i$ with probability $1-\alpha$. 

\begin{lemma}\label{lem:agg1}
Suppose SP $i$ has access to $B_i$ units of proprietary spectrum, $W_i$ units 
of licensed shared spectrum and $W_0$ units of open access spectrum; 
let $x_i$, $w_{i,L}$ and $w_{i,0}$ be the amounts of traffic 
served on each respective band in equilibrium. 
This is equivalent to a model where instead of allocating
$x_i$ and $w_{i,L}$ separately, SP $i$ allocates the total traffic 
$\tilde{x}_i = x_i + w_{i,L}$ to a single band where the price is determined by 
\begin{equation}
p_i = p_{d} -  (1-\alpha)\frac{\tilde{x}_i +w_{i,0}}{B_i} - \alpha \frac{\tilde{x}_i}{B_i+ W_i}.
\label{eq:pi}
\end{equation}
\end{lemma}
\proof
Given the traffic allocations of SP $i$ as stated in the lemma, 
the resulting price in the proprietary spectrum will be 
\[
p_i = p_{d} -   (1-\alpha)\frac{x_i + w_{i,L} +w_{i,0}}{B_i} - \alpha \frac{x_{i}}{B_i} 
\]
and the price in the shared spectrum will be
\[
p_{i}^{w,L} = p_{d} -   (1-\alpha)\frac{x_i + w_{i,L} +w_{i,0}}{B_i} - \alpha \frac{w_{i,L}}{W_i}.
\]

Note that if the SP changes $x_{i}$ and $w_{i,L}$ while keeping  $\tilde{x}_i= x_i + w_{i,L}$ fixed, 
this affects $p_i$ and $p_{i}^{w,L}$ but leaves all other prices for all other SPs and bands fixed at the same values. Hence, at any equilibrium with given $\tilde{x}_i$,
the values of $x_{i}$ and $w_{i,L}$ must solve:
\begin{equation*}
\begin{split}
\max \;\;&x_ip_i + w_{i,L}p_i^{w,L}\\
\text{s.t.} \;\;&x_i + w_{i,L} = \tilde{x}_i.
\end{split}
\end{equation*}
We can replace the objective function of this optimization problem by
\[
-x_i \frac{x_i}{B_i} - w_{i,L}\frac{w_{i,L}}{W_i}
\]
since all of the other terms only depend on the sum $x_i + w_{i,L}$.
From the first order conditions for optimality, it follows that 
\[
\frac{x_i}{B_i} = \frac{w_{i,L}}{W_i}.
\]
This implies that the price charged in each of these bands must be the same.
Further, since
\[
\frac{x_i}{B_i} = \frac{w_{i,L}}{W_i} = \frac{\tilde{x}_i}{B_i + W_i} ,
\]
this price can be written as \eqref{eq:pi}.
%\begin{equation}\label{eq:pi}
%p_i = p_{i}^{w,L} =  p_{d} -   (1-\alpha)\frac{\tilde{x}_i+w_{i,0}}{B_i} - \alpha \frac{\tilde{x}_i}{B_i+ W_i}
%\end{equation}
%which is the desired result. 
\QED

Similarly, given multiple subbands of shared spectrum that are designated as
open access, those subbands can also be pooled and treated as 
a single (intermittent) open access band with the combined bandwidth.
%Following, the same argument as used to prove this lemma, it can be seen 
%that more generally if an SP has access to multiple bands of licensed shared spectrum
%and also shared , 
%those could all be pooled into a single band with the aggregate bandwidth 
%and likewise, if there were multiple bands of unlicensed spectrum, 
%those could also be pooled. 
Hence, in the following, without loss of generality, 
we will focus on the scenario with one band of licensed shared spectrum 
per SP and at most one open access band.
The next lemma shows that in the absence of open access spectrum
we can further simplify the model and represent 
each SP as though it has an equivalent amount of licensed spectrum.

\begin{lemma}\label{lem:agg2}
Suppose SP $i$ has access to $B_i$ units of proprietary spectrum and $W_i$ units of 
licensed shared spectrum, and that there is no open access spectrum ($W_0 = 0$).  
In this case, SP $i$ can be equivalently represented as an SP 
with $T_i$ units of proprietary spectrum and no other licensed spectrum, where
\begin{equation}
T_i = B_i\frac{B_i + W_i}{B_i + (1-\alpha) W_i}.
\label{Ti}
\end{equation}
\end{lemma}

This follows from noting that when $w_{i,0} = 0$, 
the expression for $p_i$ in (\ref{eq:pi}) can equivalently be written as 
\[
p_i =  p_{d} -  \frac{\tilde{x}_i}{T_i}, 
\]
where $T_i$ is given by \eqref{Ti}.
Of course, $T_i$ is an increasing function of reliability $\alpha$
with minimum value $B_i$ (proprietary bandwidth) when $\alpha =0$, and maximum value
$B_i + W_i$ when $\alpha =1$. 

\subsection{Reliability versus Amount of Shared Bandwidth}
\begin{figure}[htbp]
\begin{center}
%\hspace{-0.5in}
\includegraphics[height=2.8in]{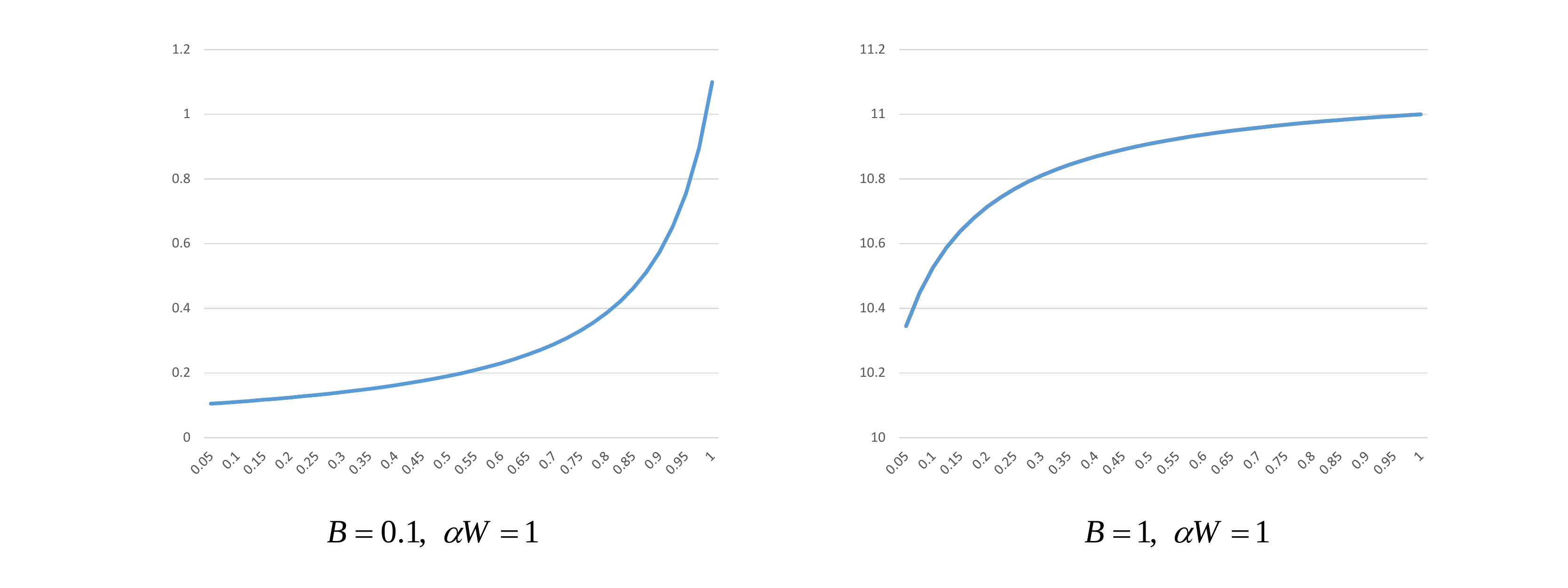}
\vspace{-0.3in}
\caption{Plots of the equivalent bandwidth $T_i$ in \eqref{Ti} versus $\alpha$
with fixed average bandwidth $\alpha W$.}
\label{fig:T}
\end{center}
\end{figure}

In practice, there may be some flexibility in determining the availability 
of the shared bandwidth, $\alpha$. That leads to a trade-off 
between $W$ and $\alpha$. Consider the case $W_i = W$, in which all bandwidth 
is allocated to SP $i$. We ask whether SP $i$ would prefer $W$ units of bandwidth 
available with probability $\alpha$, or $\alpha W$ units of non-intermittent bandwidth.
In other words, is it better to have a smaller amount of bandwidth always available, 
or a larger amount with intermittent availability, fixing the average?
Since $B_1 + \alpha W > T_1$,
the SP would prefer the smaller amount of certain bandwidth.
Fig. \ref{fig:T} shows plots of $T_i$ versus $\alpha$
with fixed $\alpha W_i$, the average amount of shared bandwidth.
The plots show that $T_i$ is monotonically increasing with $\alpha$, 
which implies that an SP always prefers higher reliability with
less bandwidth.
%given the choice between higher reliability $\alpha$
%and more bandwidth $W_i$, an SP would prefer higher reliability,
%since that produces more equivalent bandwidth $T_i$.
%In the former case, the intermittent spectrum is never available and so $W_i$ does not change a SP's latency. In the later case, the intermittent spectrum is always available and so it is equivalent to having a single band with the aggregate bandwidth.
Furthermore, the left plot shows that the variation in $T_i$ with $\alpha$
can be substantial. That corresponds to the scenario in which $W_i=1$
and $B=0.1$, so that the shared band greatly increases the amount of spectrum
potentially available. The knee of the curve, however, occurs when $\alpha > 0.7$,
indicating that the band must be relatively reliable in order to provide
a significant enhancement of available spectrum. In contrast, the variation
shown in the right plot is much smaller since $W << B$.

\subsection{Welfare Measures}

We will focus on two basic welfare measures, consumer surplus and social welfare.
The consumer surplus associated with an equilibrium allocation 
is the difference between the amount the customers receiving service 
would be willing to pay and the total cost they incur. 
Since all customers incur the same cost $p_d$, it follows that the consumer surplus is given by
\begin{equation}
 \label{eq:cs}
CS(z) = \int_0^z (P(y) - P(z)) \,dy = \frac{z^2}{2}. \\ 
\end{equation}
where $z$ denotes the total number of customers served over all bands. 
Note that $CS(z)$ is a strictly increasing function of $z$
so that to compare the consumer welfare of different equilibria 
we need only compare the number of customers served.
The social welfare of an equilibrium is the sum of the consumer surplus 
and the total revenue earned by all SPs.

%\footnote{Here we focused on a provider having one band of proprietary spectrum and one band of licensed secondary spectrum but the same argument extends directly to a model where the SPs could have multiple bands of proprietary and/or licensed secondary spectrum showing in such cases that only the aggregate bandwidth matters and so there is no loss in generality in assuming at most one band of each.}

%We next begin with an analysis of the case there are only $N=2$ service providers.

%The case $W = 0$ represents the status quo. We consider the following two approaches for the government to release $W>0$ units of new bandwidth  (which is intermittently available with probability $\alpha$). First, by dividing it up  among the SPs so that each SP receives additional proprietary bandwidth that is intermittent but exclusive to it (when not used by the government). Call this the licensed secondary regime. Second, it offers additional bandwidth as unlicensed, subject to pre-emption. Call this the unlicensed secondary regime.  Our goal is to understand the implications of each. We begin with an analysis of the licensed secondary model.

\section{Two Service Providers}\label{sec:TwoSPs}
We start with the scenario in which there are two competing SP.
This allows an illustration of the basic properties of the model.
We then consider scenarios with more than two SPs in the subsequent section.

\subsection{Shared Licensed Access}
\label{twolic}

We first examine the scenario in which all of the shared bandwidth is available for licensed access.
%under the licensed secondary regime. 
From Lemma~\ref{lem:agg2}, we can view each provider $i$ as having $T_i$ units 
of proprietary spectrum, which includes its portion of the shared spectrum.
For $N=2$, the conditions for Cournot competition reduce to:
\begin{align*}
p_d& = 1- x_1 -x_2 \\
p_1 & = p_d-\frac{x_1}{T_1}=1- x_1 \left(1+\frac{1}{T_1} \right) - x_2 \\
p_2 & = p_d-\frac{x_1}{T_2}=1- x_2 \left(1+\frac{1}{T_2} \right) - x_1 
\end{align*}
where the SPs choose $x_1$ and $x_2$, respectively. SP $i$'s revenue 
is given by $R_i= p_i x_i$, which from these relations is a quadratic 
funciton of $x_i$.

The next result characterizes the unique Nash equilibrium 
for this setting.
%\footnote{In fact, in Theorem \ref{thm:existence} 
We show in Theorem \ref{thm:existence} that the underlying game is a potential game.
From Lemma~\ref{lem:agg2}, the model reduces
to an equivalent model with no intermittent spectrum, which corresponds 
to a special case of the model studied in \cite{perakis2014efficiency}.

\begin{theorem}
\label{thm3.1}
There is a  unique Nash equilibrium  given by
\begin{align*}
x_1^* & = \frac{T_1T_2 + 2 T_1}{3T_1T_2+4+4(T_1+T_2)} \\
x_2^* & = \frac{T_1T_2 + 2 T_2}{3T_1T_2+4+4(T_1+T_2)}
\end{align*}
with the equilibrium prices given by
\begin{align*}
p_1^* & = \frac{T_1T_2+2T_1+T_2+2}{3T_1T_2+4+4(T_1+T_2)} \\
p_2^* & = \frac{T_1T_2+2T_2+T_1+2}{3T_1T_2+4+4(T_1+T_2)}.
\end{align*}
\end{theorem}

This theorem enables us to deduce the following comparitive statics.

\begin{theorem}\label{status-rev}
Let $R_i(T_1, T_2)$ be the equilibrium revenue of SP $i \in \{1,2\}$ given that each SP $i$ has $T_i$ units of equivalent proprietary spectrum.
\begin{enumerate}
\item $R_1(T_1, T_2)$ is strictly increasing and concave in $T_1$ holding $T_2$ fixed. 
\item $\frac{\partial R_1}{\partial T_2} < 0$ .
\item If $T_1 > T_2$, then $\frac{\partial R_1}{\partial T_1} > \frac{\partial R_2}{\partial T_2}$.
\end{enumerate}
\end{theorem}
Theorem \ref{status-rev} has two immediate implications. First, unsurprisingly, each SP would prefer to have larger amounts of the equivalent shared bandwidth than not, 
other parameters held fixed. Second, an increase in the equivalent shared bandwidth 
of one's rival results in a decrease in one's own revenue.   
Interestingly, because $T_i$ increases with $B_i$, the marginal value of 
additional equivalent licensed bandwidth is {\em larger} for the SP 
with the larger initial amount of bandwidth.

From \eqref{eq:cs}
%Given the inverse demand relationship \eqref{Py}, 
the consumer surplus 
%for total traffic carried: $z=x^*_1 + x^*_2 $ 
is given by
\begin{align}
CS(z) = 
%\int_0^z(1-y)dy-(1-z)z = \frac{z^2}{2} = 
\frac{(x^*_1 + x^*_2)^2}{2},
\label{cs}
\end{align}
and from Theorem \ref{thm3.1} we have
\begin{align}
\nonumber
x_1^* + x_2^* & = \frac{T_1T_2 + 2 T_1}{3T_1T_2+4+4(T_1+T_2)} +
 \frac{T_1T_2 + 2 T_2}{3T_1T_2+4+4(T_1+T_2)} \\
& = \frac{2T_1T_2 + 2 (T_1+T_2)}{3T_1T_2+4+4(T_1+T_2)} .
\label{x1+x2}
\end{align}
 %If we normalize $T_1 + T_2 = 1$, then, 
 %$$x_1^* + x_2^* = \frac{2T_1T_2 + 2 }{3T_1T_2+8}.$$
Referring to the expression for $T_i$ in \eqref{Ti},
%Recall that $$T_i = B_i\frac{B_i + W_i}{B_i + (1-\alpha)W_i}\,\, \forall i = 1, 2.$$
consumer surplus is therefore a non-linear function of $W_1, W_2$. 
This means that for some parameter settings consumer surplus will {\em not} 
be maximized by setting $W_1 = W$ or $W_2 = W$. 
To understand the implication of this suppose the incumbent 
decides to allocate all $W$ units of new bandwidth by auction 
to the highest bidder. The resulting allocation need not maximize consumer surplus.
If $B_1 = B_2$, then one can show that the new spectrum should 
be divided equally between the two SPs to maximize consumer welfare.  
In general, the allocation that maximizes consumer surplus will
make the $T_i$ of the SP with the larger $B_i$ larger than the $T_i$ 
of the other SP (though the smaller SP may still get a larger amount of $W_i$). 
It is also the case that the SP with the larger $B_i$ benefits 
more from an increase in spectrum that is intermittent. (The smaller $\alpha$ is,
the greater the difference in benefit.)
This is because the larger SP is better able to absorb the fluctuations. 

\subsection{Shared Open Access}
\label{twoshared}
%{\color{red} Changing $B\Rightarrow B/2$ for consistency and ease of comparison with many providers case.}
We now assume that each SP has its own proprietary bandwidth and that
the additional shared bandwidth $W$ is open access.
%incumbent provides additional bandwidth $W$ to be shared.
%Here we examine the consequences of allocating additional bandwidth, $W$, between two service providers under the unlicensed secondary regime. 
We begin with the symmetric case where each SP has the same amount 
of proprietary bandwidth, i.e., $B_1 = B_2 = B/2$. 
\begin{theorem}
\label{thm3.4}
When $B_1 = B_2 = B/2$ the unique Nash equilibrium is given by the quantities
\begin{align*}
%x_i & = \frac{\frac{3}{W}}{\frac{9}{W}+\frac{6}{B}+\frac{6}{B W}+\frac{4 (1-\alpha)}{B^2}} =  \frac{3B^2}{9B^2+6BW+6B+4 (1-\alpha)W} \\
%w_i  & = \frac{\frac{2}{B}}{\frac{9}{W}+\frac{6}{B}+\frac{6}{B W}+\frac{4 (1-\alpha)}{B^2}} = \frac{2BW}{9B^2+6BW+6B+4 (1-\alpha)W}.
x_i^* & = \frac{\frac{3}{W}}{\frac{9}{W}+\frac{12}{B}+\frac{12}{B W}+\frac{16 (1-\alpha)}{B^2}} =  \frac{3B^2}{9B^2+12 BW+12 B+16 (1-\alpha)W} \\
w_i^*  & = \frac{\frac{4}{B}}{\frac{9}{W}+\frac{12}{B}+\frac{12}{B W}+\frac{16 (1-\alpha)}{B^2}} = \frac{4BW}{9B^2+12 BW+12 B+16 (1-\alpha)W}.
\end{align*}
for $i=1,2$. 
\end{theorem}

In particular, {\em both} SPs make use of the unlicensed bandwidth.
Direct computation using the preceding quantities shows that 
both SP revenue and consumer surplus increase with the following parameter variations:
\begin{enumerate}
\item $B$ increases holding $W$ and $\alpha$ fixed;
\item $W$ increases holding $B$ and $\alpha$ fixed;
\item $\alpha$ increases holding $B$ and $W$ fixed.
\end{enumerate}

We remark that for the analogous Bertrand model
considered in \cite{nguyen2014cost}, the SP revenue generally {\em decreases}
as $W$ increases. This is due to the shift in customers to the open access
band, where the equilibrium price is zero. In contrast, for the Cournot model
the corresponding shift in traffic generally lowers the price, but that 
is offset by an increase in the number of customers served.

\Xomit{
The next result shows that the magnitude of the increase in consumer surplus from the introduction of additional shared bandwidth depends on the amount offered relative to the amount of propretary bandwidth.
\begin{theorem}
\label{thm3.5}
There exists a $\tau$ depending on $\alpha$ such that when $B < \tau W$, the increase in consumer surplus is smaller under the unlicensed secondary regime than dividing it equally between the two service providers. When $B > \tau W$, offering  additional bandwidth under the unlicensed secondary regime increases consumer surplus more than dividing it equally between the two providers.
\end{theorem} 

 {\color{red} add intuition and contrast with the large $n$ case from earlier.}
 }

%From the previous section **check** it follows that there is a unique equilibrium.  It is easy to argue that prices being $0$ at equilibrium can only occur  if the quantity is also zero: if not, then reducing the quantity by $\epsilon$ increases the price and revenue. We will concentrate on the two provider scenario and analyze asymmmetric equilibrium in further detail.
We next consider the {\em asymmetric} scenario in which $B_1 \neq B_2$.
As $B_1$ increases relative to $B_2$ we obtain the following result.
%With extreme asymmetry we have the following that is related to proposition \ref{prop:asym}, later in this paper.
%special case of Proposition \ref{prop2}:
\begin{theorem} 
\label{thm:asym}
%For $\beta = 1$, i.e., all the new bandwidth is shared,  there is a unique equilibrium with 
%$w_1 = 0$, $x_1 > 0$ and $x_2 > w_2 > 0$ if and only if
%$B_1 \geq 2 W + 4(1-\alpha) \frac{W}{B_{2}}+2 B_{2} +2$. 

Suppose that a fraction $\beta W$ of the shared bandwidth is provisioned as open access, 
and the remainder is partitioned as $W_i$, $i=1,2$, and allocated as licensed 
bandwidth to the respective SPs, where $\beta \in [0,1]$ and $W_1+W_2=(1-\beta)W$.
Then, there is a unique equilibrium with $w_1^* = 0$, $x_1^* > 0$ and $x_2^* > w_2^* > 0$ if and only if
\begin{align}\label{eq:asymN2}
B_1+W_1 + 2(1-\alpha) \frac{W_1}{B_1} \geq 2 (\beta W + W_2) + 2 B_2 + 2 + 4 (1-\alpha) \frac{\beta W + W_2}{B_2}.
\end{align}
For $\beta=1$, all the shared bandwidth is open access
and condition \eqref{eq:asymN2} simplifies to 
$B_1 \geq 2 W + 4(1-\alpha) \frac{W}{B_{2}}+2 B_{2} +2$.
\end{theorem}
%{\color{red} Need to check the proof as I suspect that the term $4 (1-\alpha) \tfrac{\beta W + W_2}{B_2}$ in \eqref{eq:asymN2} is actually $2 (1-\alpha) \tfrac{2 \beta W + W_2}{B_2}$.}

%The proof is given in Appendix \ref{app:asym} and consists of showing that the condition corresponds to an equilibrium that maximizes the potential function \eqref{eq:potfn} given in the next section.
Hence, if an SP has an amount of proprietary bandwidth that greatly exceeds that
held by the other SP, there is a range of $W$ for which
it will {\em not} make use of the shared bandwidth,
leaving it for the smaller SP.
%In other words, only the smaller provider makes use 
%of the shared bandwidth. 
Interestingly, as $\alpha$ decreases, i.e., the shared band is more likely to be pre-empted, 
the `larger' SP is more likely to use it. 
This is because its proprietary bandwidth makes it better able 
to handle the traffic in the event of pre-emption.

The proof is given as part of Appendix \ref{sec:potential} (see section \ref{app:structure}).
There it is also shown that conditioned on the shared spectrum being available,
\begin{align*}
\frac{x_i^*}{B_i+W_i} \leq \frac{w_i^*+\frac{w_{-i}^*}{2}}{\beta W} \leq \frac{w_1^*+w_{2}^*}{\beta W}, 
\end{align*}
for $i\in \{1,2\}$, and where $-i$ denotes the other SP.
Suppose that $\beta=1$, so that $W_i=0$.
The condition then states that when the shared band is available,
the congestion in the proprietary bands 
is always less than the congestion in the open access band.
If SP $i$ uses the shared band, it is shown that the first inequality is tight. 
Additionally, if in equilibrium $w_i^* > 0$ for both SPs,
then for SP $i$ the congestion in the open access band is strictly greater
than the congestion in its licensed bands by ${w_{-i}^*}/{(2W)}$.
Furthermore, if in equilibrium $w_i^*=0$, 
%SP $i\in \{1,2\}$ 
%does not use the shared band, 
then the open access band and
proprietary band for the other provider $-i$ have the {\em same} congestion level,
which is at least twice the congestion in SP $i$'s proprietary band. 
It is this additional congestion which causes SP $i$ to assign $w_i=0$,
and to use only its proprietary bands.
%Interestingly, this is the only reason for provider $i$ to not use the shared band, 
%as opposed to others such as the (corresponding) price becoming negative 
%if the traffic carried is positive, etc.
The appendix also considers $N>2$ asymmetric SPs, and gives a condition
for when all but one SP assigns traffic to the open access band.

%We also have the following property of the equilibrium bandwidth allocation.
%\begin{prop}
%At any equilibrium $x_i^*>0$ for $i=1,2$. Furthermore,
%if the condition in Theorem \ref{thm:asym} does not hold, 
%then the equilibrium is an interior point. 
%\end{prop}
%The proof is in Appendix \ref{app:asym}.
\subsection{Numerical Examples}
To illustrate the behavior of the SPs in scenarios
not covered by Theorem \ref{thm:asym} we present a series of numerical examples
showing comparative statics with different assumptions concerning
how the shared bandwidth is allocated.

\begin{figure}[htbp]
\centering
%\begin{center}
\centering
\begin{subfigure}{5.5in}
\includegraphics[width=5.5in,height=3.0in]{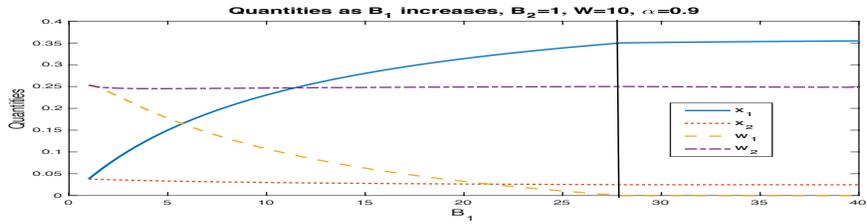}
\caption{Quantities}
\label{fig:quantities}
%\end{center}
\end{subfigure}
\begin{subfigure}{5.5in}
\centering
%\begin{center}
\includegraphics[width=5.5in,height=3.0in]{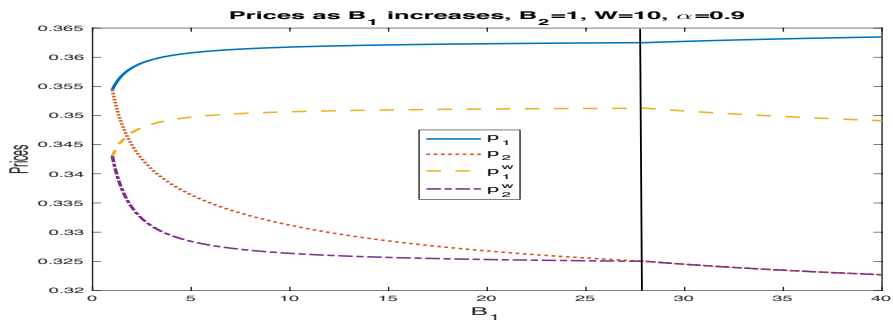}
\caption{Prices}
\label{fig:prices}
\end{subfigure}
\caption{Equilibrium quantities $x_i$, $w_i$, and prices $p_i$, $p_i^w$, $i=1,2$, 
as $B_1$ increases with $B_2=1$, $W=10$ and $\alpha=0.9$. }
%\end{center}
\label{fig:prices_quantities}
\end{figure}

\paragraph{Quantities and prices:}
Figure \ref{fig:prices_quantities} illustrates how quantities and prices
in the shared and proprietary bands change as the amount
of proprietary bandwidth for SP 1 ($B_1$) increases. 
Figure \ref{fig:quantities} shows that as $B_1$ increases,
SP 1 increases the quantity of customers $x_1$ served in its proprietary band,
and decreases its allocation of customers $w_1$ to the shared band.
SP 2 maintains nearly constant quantities in both bands.
The vertical line shows the threshold $B_1^*$ at which $w_1$ becomes zero.
For $B_1 > B_1^*$ the quantities are nearly constant with only
slight variations due to the limited competition with only two SPs.

Figure \ref{fig:prices} shows that SP 1 charges higher prices than SP 2
in both bands since it is able to provide lower latency than SP 2. 
For $B_1 < B_1^*$, SP 1's prices increase, due to decreasing latency, 
whereas SP 2's prices decrease to maintain its quantity of customers.
For $B_1 > B_1^*$, $p_1$ increases slowly, since latency in that band
continues to decrease, whereas the remaining prices decrease
to maintain the nearly constant quantities shown in Figure \ref{fig:quantities}.

\begin{figure}[htbp]
\centering
\begin{subfigure}{5.5in}
%\begin{center}
\centering
\includegraphics[height=2.5in,width=5.5in]{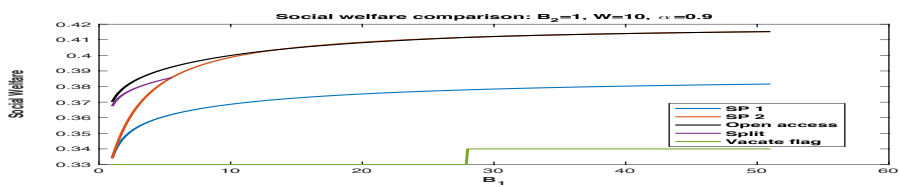}
\caption{$B_1$ varies with $W=1$}
\label{fig:SW_B}
%\end{center}
\end{subfigure}
%{\color{red}  $K_i$ to be changed to $B_i$. Lower case $w$ should  be changed to $W$. Need explanation of graph. What do words like split mean?}
\begin{subfigure}{5.5in}
%\begin{center}
\centering
\includegraphics[height=2.5in,width=5.5in]{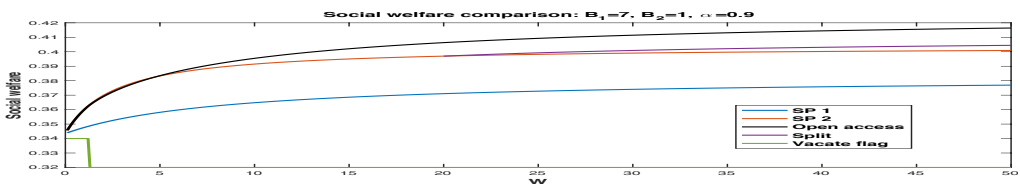}
\caption{$W$ varies with $B_1=7$}
\label{fig:SW_W}
\end{subfigure}
%\end{center}
\caption{Comparison of social welfare for various schemes for allocating the shared bandwidth $W$; 
$B_2=1$ and $\alpha=0.9$.}
\label{fig:SW}
\end{figure}

\paragraph{Social welfare:}
Figure \ref{fig:SW} shows social welfare achieved by four different schemes 
for allocating the shared spectrum. These schemes are motivated by the discussion 
following Theorem \ref{status-rev}, which considers the outcome of a winner-take-all
auction of the shared band. The label ``SP 1" in Figure \ref{fig:SW} indicates
all of the shared spectrum is allocated to SP 1, which always possesses 
the greater amount of proprietary spectrum. Similarly,
the label ``SP 2" allocates all of the shared spectrum to SP 2.
We compare the social welfare for these outcomes with that obtained
by allocating the shared spectrum as open access, labeled ``Open access".
%{\color{red} should change label}
Finally, the label ``Split" allocates the shared spectrum 
to equalize the equivalent always-available bandwidths $T_1$ and $T_2$, 
if possible, or otherwise allocates all of the shared bandwidth to SP 2.

Figure \ref{fig:SW_B} depicts how social welfare changes 
as a function of $B_1\geq B_2=1$ (with $\alpha=0.9$ and $W=10$). 
Assigning $W$ to the smaller provider SP 2 always achieves higher social welfare
since this enables SP 2 to compete more effectively with SP 1.
However, as implied by Theorem \ref{status-rev}, SP 1 has an incentive
to bid a higher amount for $W$ than SP 2. The resulting loss in social welfare
is indicated in the figure.
This analysis suggests that an auction for $W$ should be enhanced 
to contain more options.\footnote{A related auction 
for the allocation of a congestible resource is proposed in \cite{Barrera2015}.
Our setting is more difficult because there
are two externalities to be managed: congestion and downstream competition.}
An example is provided in Appendix \ref{app:bids},
which considers the scenario in which the SPs can bid
for the shared spectrum with the following options:
it is licensed entirely to SP 1 or SP 2, or it is 
shared as open access. The example shows that both
SPs may prefer that the shared bandwidth be open access
rather than licensed.

For the schemes considered in Figure \ref{fig:SW}, 
open access sharing yields the highest social welfare except 
for a small region where the scheme ``SP 2" does marginally better. 
The ``Vacate flag" indicates the values 
of $B_1$ for which SP 1 does not use the shared spectrum, 
so that it is effectively allocated to SP 2. Hence in that region the social welfare
for open access coincides with that for scheme SP 2.
%Allocating $W$ to SP 1 provides the lowest social welfare owing to the 
%negative externality imposed on the smaller service provider. 

When it is possible to set $T_1=T_2$, the ``Split" scheme 
achieves a higher social welfare than either scheme SP 1 or SP 2. 
This indicates that among schemes that partition the shared spectrum 
between the two SPs, there is an optimal split that lies
between the two extreme schemes SP 1 and SP 2. 
Open access sharing can be thought of as a more flexible split between the two SPs.
Figure \ref{fig:SW_W} depicts how social welfare changes as $W$ 
increases for the four allocation schemes considered. 
The relative differences observed previously also
apply in this regime. However, for open access sharing, 
as $W$ increases, in equilibrium, SP 1 always uses the shared spectrum.
%{\color{red}  $K_1$ should be changed to $B_1$. Need explanation of graph. What do words like split mean?}

\section{Many Symmetric Service Providers}\label{sec:manySPs}
\label{symprov}
We now examine the scenario with an arbitrary number of SPs $N$,
where the SPs are symmetric. That is, they each have the same amount
of licensed bandwidth (including licensed shared bandwidth). 
Specifically, the shared band is split into an open access part with
bandwidth $\beta W$ and a licensed, or proprietary part with bandwidth $(1-\beta)W$ 
for a fixed $\beta \in [0,1]$. Each SP has its own proprietary bandwidth $B/N$,
and the licensed part of the shared band is split equally
among the SPs. The shared band therefore adds $(1-\beta)W/N$ units of licensed bandwidth
to each SP with availability $\alpha$.
We will compare the total welfare, total revenue, and consumer surplus
when the shared band is allocated as licensed versus open access.
Analytical results are presented for $N \to \infty$, reflecting perfect competition.
%In this setting the unique equilibrium will be symmetric.

%case in which the number $N$ of competing SPs
%within a fixed allocation of bandwidth $B$ becomes large, reflecting perfect competition.
%Here, we fix the amount of bandwidth and increase $n$. This captures the idea that the unlicensed secondary regime will encourage a flood of new entrants. 
%We first assume a symmetric scenario in which each SP has $B/N$
%units of licensed, or proprietary bandwidth, and subsequently consider an asymmetric scenario.
%As before, the additional shared band has bandwidth $W$, and is
%intermittently available with probability $\alpha$.

%\subsection{Symmetric Providers}

Let $(\bar{x}(N),\bar{w}(N))$ denote the symmetric equilibrium allocation,
{\em i.e.}, $\bar{x}(N)$ and $\bar{w}(N)$ are the same for all SPs.
Here $\bar{x}$ and $\bar{w}$ refer, respectively, to the quantities
allocated to the licensed bands, including the licensed part of the shared band,
and the open access band.

%$nW_1$ units as licensed and $W_2$ units as unlicensed. Thus $nW_1 + W_2 = W$ and each service provider receives a grant of $W_1$ units of intermittently available bandwidth.
%let $B$ be total bandwidth of  proprietary  spectrum; $W$ be total  bandwidth of  new secondary. there are $n$ SPs.  Consider scenario in which  $W_1$  is  used as licensed, and to be divide among SPs; $W_2$ is used as unlicensed.

\begin{lemma}\label{prop:01}
For any finite $N$ the equilibrium is symmetric and unique.
\end{lemma}

\begin{theorem} \label{corr:01}
As $N\to\infty$, the limiting equilibrium is specified by
$$(x^*,w^*)= \lim_{N\rightarrow \infty} (N \bar{x}(N), N\bar{w}(N)),$$ where
\begin{equation} \label{eq:xsws}
\begin{split}
x^*&= \frac{\big(B+(1-\beta) W\big) B}{\big(B+2 (1-\alpha) \big) \big( B + W (1+\beta) \big)+2 B \alpha}, \\ 
w^*&= \frac{2 \beta W B}{\big(B+2 (1-\alpha) \big) \big( B + W (1+\beta) \big)+2 B \alpha},
\end{split}
\end{equation}
and the limiting prices in the licensed and open access bands are given by
 \begin{equation} \label{eq:pandpw}
 \begin{split}
  p&=\frac{(1-\alpha) \big( B + W (1+\beta) \big)+B \alpha}{\big(B+2 (1-\alpha) \big) \big( B + W (1+\beta) \big)+2 B \alpha}, \\
 p^w & =  \frac{(1-\alpha) \big( B + W (1+\beta) \big)}{\big(B+2 (1-\alpha) \big) \big( B + W (1+\beta) \big)+2 B \alpha}.
 \end{split}
  \end{equation}
\end{theorem}
The proof is given in Appendix \ref{app:manyusers}. There the expressions
for $\bar{x}(N)$, $\bar{w}(N)$ are given for arbitrary $N$.

Theorem \ref{corr:01} has the following implications.
\begin{enumerate}
\item From \eqref{eq:pandpw} $p > p^w$ for all $\alpha>0$.
Also, from \eqref{eq:xsws}, the congestion, or load in the open access band
(users/total bandwidth) is
\begin{equation}
\label{eq:xandw}
\frac{w^*}{\beta W} = 2 \frac{x^*}{B+(1-\beta) W} .
\end{equation} 
Therefore, as $N \to \infty$, the congestion in the open access band 
is {\em twice} the congestion in the licensed/proprietary band,
and the announced price for the open access band is lower than that
for the licensed band. That is because the SPs have an incentive 
to shift traffic into the open access band, and then charge
a higher price for the lower latency experienced in the licensed band.
More generally, for arbitrary $N$, the expressions in the appendix
show that the congestion in the open access band is $2N/(N+1)$ times 
the congestion in the proprietary bands.
\item Unlike the classical Cournot model of competition, 
the prices do not converge to zero as the number of competing agents becomes large.
This is due to the tradeoff between announced price and congestion cost.
\item In the special case where the shared band is always available, %$B=0$ and 
$\alpha=1$ and
\begin{align}
\begin{split}
%x^*&=\frac{(1-\beta)W}{W(1+\beta)+2},\; w^*=\frac{\beta W}{W(1+\beta)+2}, \; x^*+w^*=\frac{W(1+\beta)}{W(1+\beta)+2}, \\
p&=\frac{1}{W(1+\beta)+B+2}, \; p^w=0.
\end{split}
\end{align}
That is, the price for the open access band is zero. 
This is analogous to the equilibrium with Bertrand (price)
competition, derived in \cite{nguyen2014cost}. There it is also observed
that the price of the open access band is zero, although
here that occurs only when there are sufficiently many SPs.
\end{enumerate}

%{\thanhedit
%There are several takeways from this proposition:
%\begin{itemize}
%\item unlike standard cournot competition, here prices do not converge to 0.\\ this is important: should say something more.... because of congestion? because of  intermittent spectrum?
%\item According to \eqref{eq:pandpw} price at licensed bands is higher than at unlicensed. WHY?
 % \item 
%From  \eqref{eq:xsws} We have 
%\begin{equation}
%\label{eq:xandw}
% \frac{x^*}{B+(1-\beta) W}=\frac{w^*}{2\beta W}. 
%\end{equation} 
%This  has special meaning: expected congestion in licensed spectrum is half expected congestion at  unlicensed? WHY? because strategy of common.  
%\end{itemize}
%}
\begin{theorem} \label{limit1}
As $N \rightarrow \infty$, consumer surplus is maximized when $\beta=1$.
However, total revenue and social welfare are maximized when $\beta=0$.
 %(I think i got this calculation. Need to explain the intuition..)
\end{theorem}
From \eqref{eq:xsws}, the total traffic carried is given by
\begin{align*}
x^*+w^*=\frac{B \big( B + W (1+\beta) \big)}{\big(B+2 (1-\alpha) \big) \big( B + W (1+\beta) \big)+2 B \alpha}.
\end{align*}
Hence the total traffic along with consumer surplus is maximized when $\beta=1$.
The rest of the proof is given in Section \ref{prooflimit1}. 

\begin{figure}[htbp]
\begin{center}
\includegraphics[height=3in,width=5.5in]{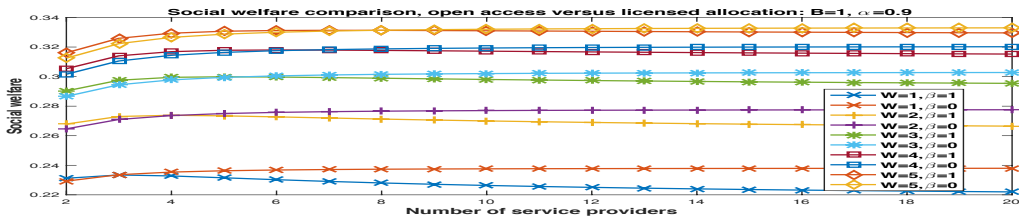}
\caption{Social welfare achieved for $\beta=0$ and $\beta=1$ 
as a function of the number of (symmetric) SPs $N$ 
for different values of $W$. The total proprietary bandwidth $B=1$ and $\alpha=0.9$}
\label{fig:SW_N}
\end{center}
\end{figure}

%\paragraph{Social welfare versus $N$:}
Figure \ref{fig:SW_N} illustrates the change in social welfare that
takes place as $N$ increases. The plots show social welfare versus $N$
for both $\beta = 0$ (all licensed) and $\beta = 1$ (all open access), 
and for different values of $W$. Focusing on the bottom two curves for $W=1$,
the curves cross when $N \leq 3$, i.e., for $N<3$, open access achieves 
higher social welfare than licensed access, and vice versa for $N>3$.
This is consistent with Theorem \ref{limit1}.
%Once $N$ exceeds 3, it flips, and social welfare is higher under $\beta = 0$ consistent with Theorem \ref{limit1}. 
Note that the corresponding crossover value of $N$ increases as $W$ increases.

\subsection{Degraded Sharing}
So far we have assumed that the latency experienced by serving traffic load $x$ 
with bandwidth $W$ is $x/W$ regardless if the traffic load comes from 
a single SP or multiple SPs. In practice, because less coordination is expected
among SPs in the open access band, it could be that
the latency experienced for a given load in the open access band
is {\em greater} than if the same load were served by a single SP.
%That due to there bing less coordination among the firms using the spectrum.
That would make licensed access more attractive and with enough degradation, 
the conclusion of Theorem~\ref{limit1} that open access maximizes consumer surplus 
may no longer hold. In this section we examine this possibility. 

We model the degradation associated with open access by introducing 
a degradation factor $d \leq 1$ and assume that when traffic load $x$ 
is served with open access bandwidth $W$, that the congestion cost is 
\[
\ell_W(x) = \frac{x}{dW}.
\]
In other words, the ``effective bandwidth" seen by the users 
of an open access band is $dW < W$. (Equivalently, the latency increases by $1/d$.)
%The next result shows that such degradation needs to be significant 
%in order to change the conclusion that open access provides higher consumer welfare.

\begin{lemma}\label{4.4}
In the symmetric model with $N\geq 2$ firms, 
if $d>\frac{N+1}{2N}$, allocating all of the shared spectrum 
as open access maximizes consumer welfare, 
whereas if $d<\frac{N+1}{2N}$, allocating all of the shared spectrum 
as licensed maximizes consumer welfare.
\end{lemma}

The proof is given in Appendix \ref{app:CS}, and is a consequence
of the first property following Theorem \ref{corr:01}.
Note that the threshold $\frac{N+1}{2N}$ does not
depend on $\alpha$, $W$ or $B$. This threshold is $3/4$ when $N=2$ 
and decreases to $1/2$ as $N$ becomes large.
Hence for large $N$, unless the latency in the open access spectrum
is more than twice that for licensed use due to lack of coordination, 
given the same load,
open access still achieves larger consumer surplus.

\subsection{Latency, price, and social welfare}
\begin{figure}
\centering
\begin{subfigure}{5.5in}
  \centering
  \includegraphics[width=5.5in]{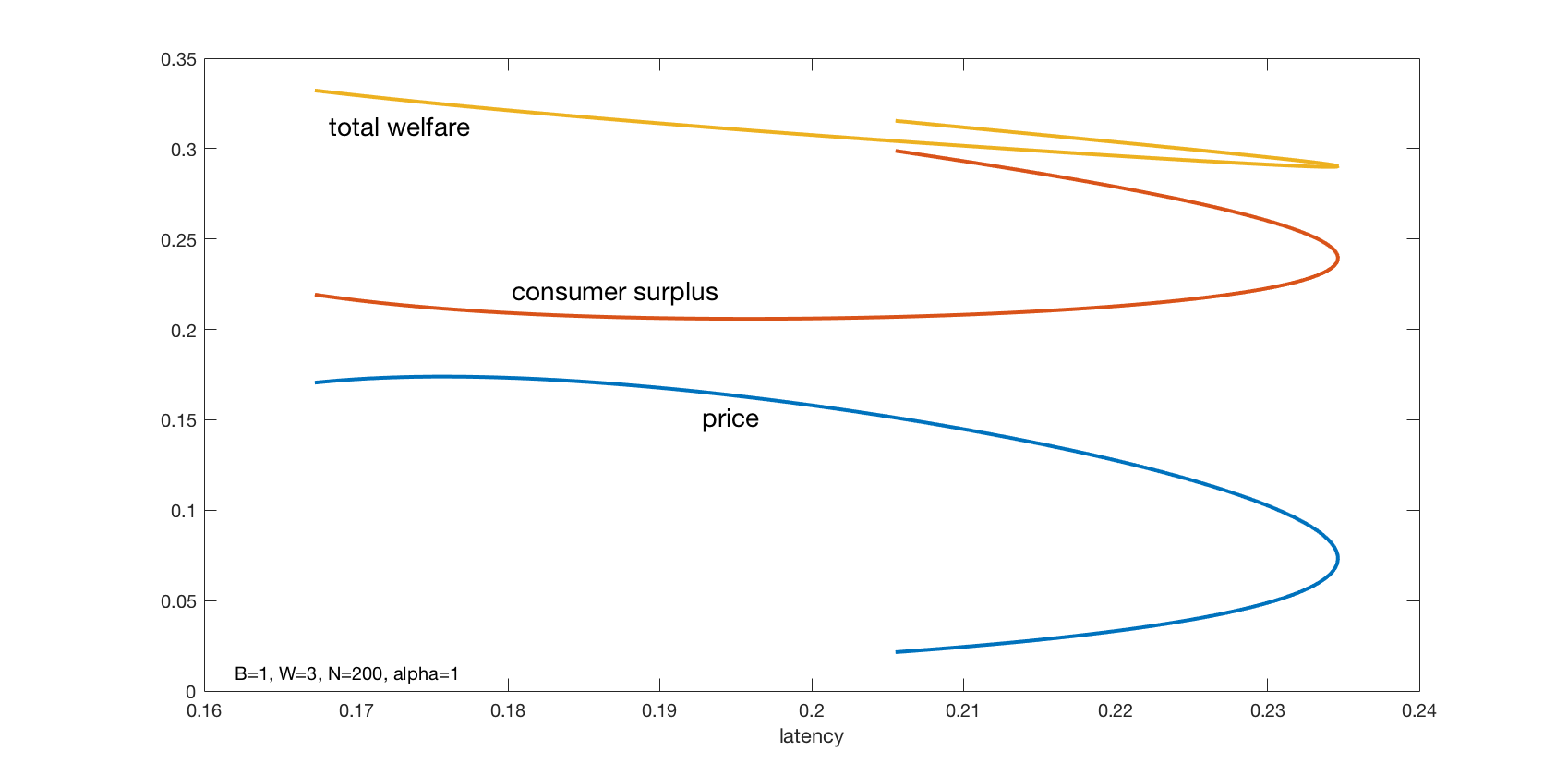}
  \caption{$N=200$}
  \label{price_lat1}
\end{subfigure}\\
\begin{subfigure}{5.5in}
  \centering
  \includegraphics[width=5.5in]{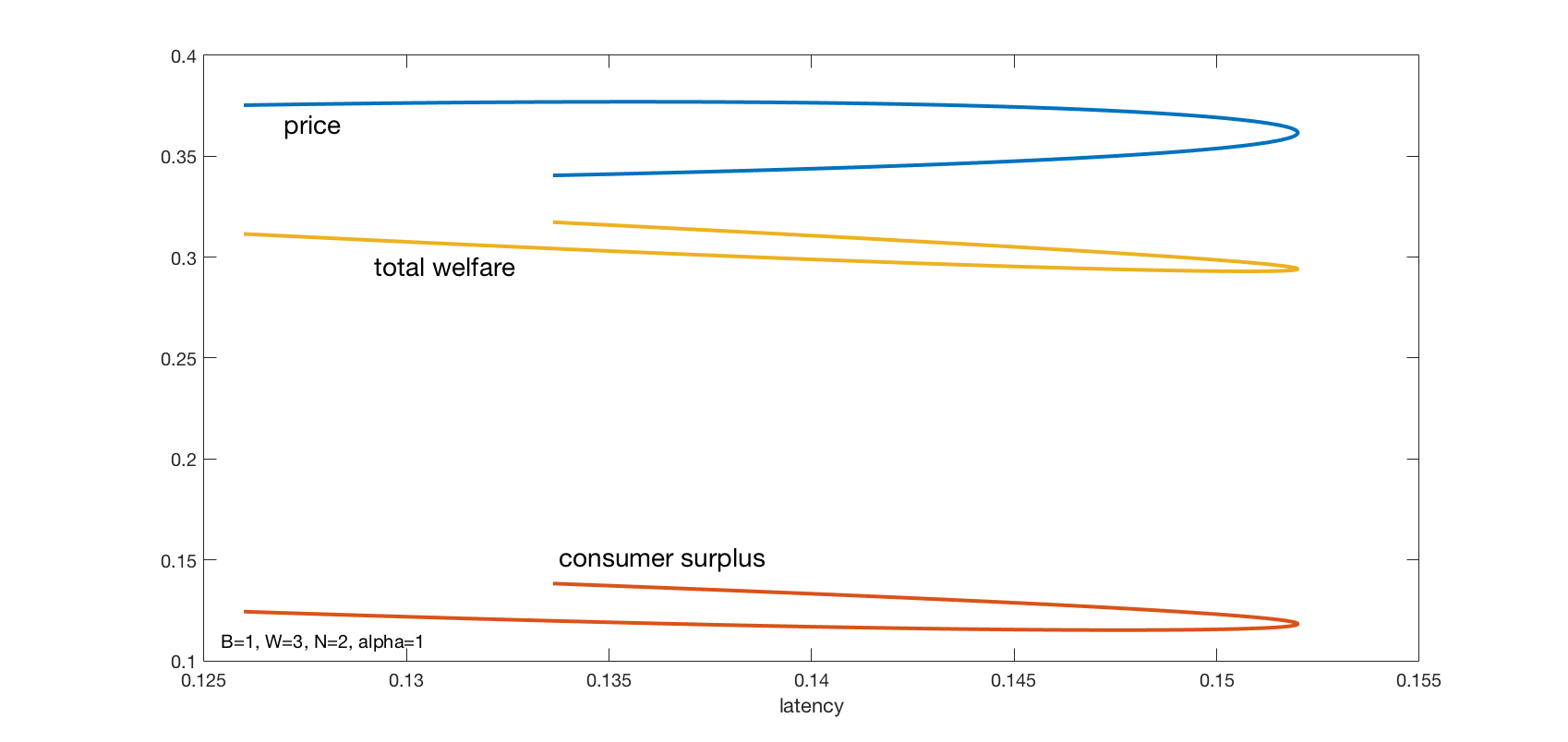}
  \caption{$N=2$}
  \label{price_lat2}
\end{subfigure}
\caption{Parametric plots of average price, consumer surplus, and total welfare versus
average latency as the fraction of shared bandwidth $\beta$ increases from zero to one.}
\label{price_lat}
\end{figure}

To gain further insight into the effects of open access bandwidth on latency and price,
Figures \ref{price_lat1} and \ref{price_lat2} show parametric plots of average price,
consumer surplus, and total welfare versus average latency
as the fraction of open access bandwidth $\beta$ increases from zero to one.
The average price is given by Lemma \ref{lem:agg1},
%\eqref{avg_p}
 and the average latency 
is similarly
\begin{equation}
\bar{\ell}(N) = \frac{\bar{x}(N)}{\bar{x}(N)+\bar{w}(N)} \bar{\ell}_p(N) + \frac{\bar{w}(N)}{\bar{x}(N)+\bar{w}(N)} \bar{\ell}_{w}(N)
\end{equation}
where
\begin{equation}
\bar{\ell}_p(N) =\alpha \frac{N \bar{x}(N)}{B+(1-\beta)W} + (1-\alpha) \frac{N(\bar{x}(N) + \bar{w}(N))}{B}
\end{equation}
and
\begin{equation}
\bar{\ell}_w(N) = \alpha \frac{N \bar{w}(N)}{\beta W} + (1-\alpha) \frac{N(\bar{x}(N)+\bar{w}(N))}{B}
\end{equation}
are the latencies associated with the proprietary and shared bands,
respectively. The figure shows plots for $N=200$ and $N=2$,
and $\alpha=1$.

No sharing ($\beta=0$) corresponds to the lowest latency on each curve
(left-most point),
and as $\beta$ increases from zero to one, the latency increases to the highest
value (right-most point), and then subsequently decreases to the final point 
corresponding to full sharing ($\beta=1$). Focusing on Fig. \ref{price_lat1},
as $\beta$ increases from zero, the average price increases slightly 
as latency increases. This is because
when $\beta W$ is small, the shift in load from the proprietary to shared band
congests the shared band, increasing both average latency and price.
In this region the consumer surplus and total welfare decrease.
As $\beta$ increases further,
the SPs lower the price to continue to shift load
to the shared band, and the average latency continues to increase.
In this region the consumer surplus {\em increases} while the total welfare
continues to {\em decrease} due to the decrease in SP revenue.
Finally, as $\beta$ is further increased towards one, both the price and latency fall,
and consumer surplus increases more rapidly, causing total welfare to increase.
Even so, the total welfare with full sharing is slightly below that
with no sharing, as expected from Theorem \ref{limit1}.

Comparing Figure \ref{price_lat1} with \ref{price_lat2}, the additional competition
with $N=200$ results in a lower price and higher consumer surplus. Furthermore,
the increase in open access bandwidth has a more pronounced effect on the quantities shown.
Further examples with $\alpha<1$ show consistent trends, but with less variation
with latency due to the diminished benefit of adding the shared bandwidth.
%The case of asymmetric SPs is harder to analyze, but calculations for a variety of special cases suggest that it is still the case that consumer surplus is maximized when $\beta=1$.  Plots of how consumer surplus varies with $\beta$ when the SPs have different amounts of propietary bandwidth are shown in Figures ???? and ????.

%Below we show that in the shared regime, SPs with a sufficiently large amount of proprietary bandwidth will never use the additional bandwidth provided by the Government.

\subsection{Effects of Increasing $W$}
\begin{figure}
  \centering
  \includegraphics[width=5.5in]{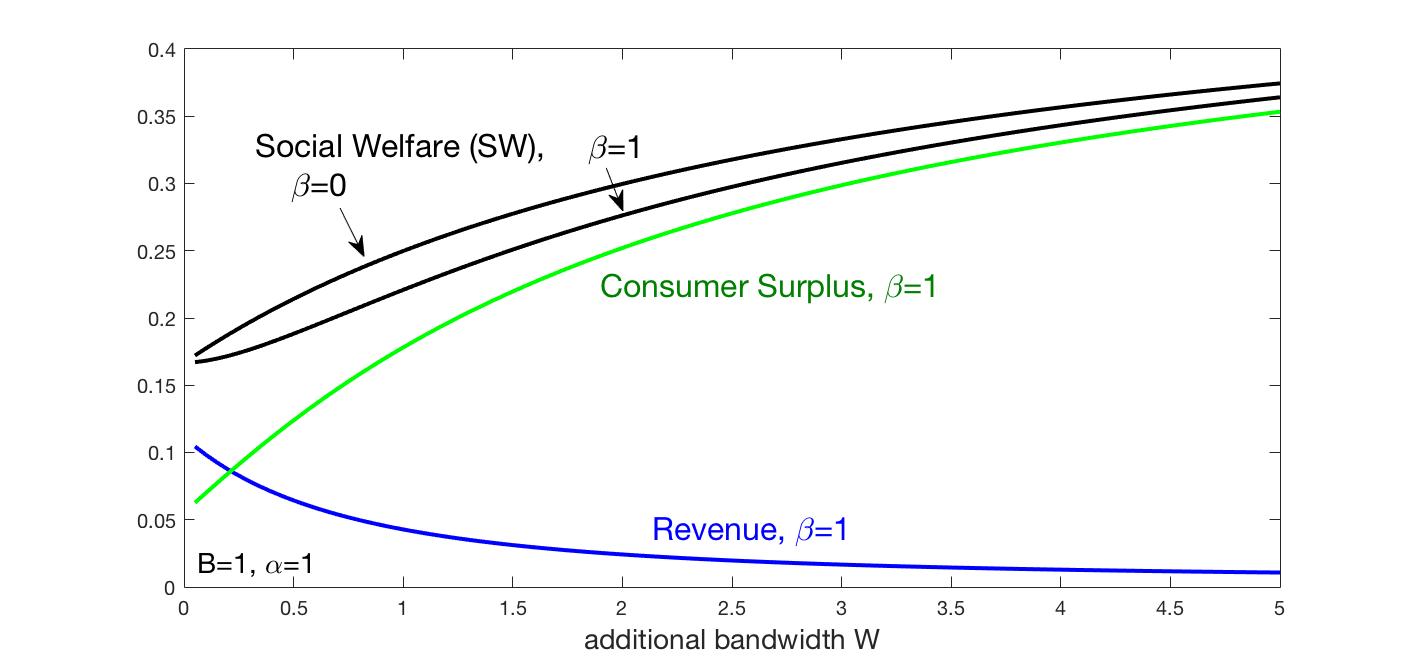}
\caption{Total welfare versus additional bandwidth $W$ with a large
number of SPs when the shared band
is open access ($\beta=1$) and proprietary ($\beta=0$). Revenue and consumer surplus
are also shown for $\beta=1$.}
\label{sw_W}
\end{figure}

%\paragraph{Social welfare versus $W$ (large $N$):}
Fig. \ref{sw_W} illustrates the effect of increasing $W$
on social welfare, consumer surplus, and revenue
for large $N$.
%holding $N$ fixed at 200.
Social welfare is shown for the cases where the shared bandwidth
is entirely open access ($\beta=1$) and proprietary ($\beta=0$).
%The gap between these curves becomes narrower as $W$ increases,
%and they eventually converge.
Both curves are monotonically increasing, but their shape changes from
convex to concave when the shared bandwidth changes from open access to proprietary.
In particular, the slope at $W=0$ is zero when the shared band is open access,
but is positive when the shared band is proprietary.
This behavior has also been observed within the Bertrand model
of price competition  \cite{nguyen2014cost}. There,
with a small number of SPs, adding a small amount of open access
bandwidth can {\em decrease} the social welfare ({\em i.e.}, the slope
at $W=0$ can be negative). Here the additional incremental shared bandwidth
increases social welfare for smaller values of $N$ (not shown),
but the increase tends to zero as $N$ becomes large.
The curves for revenue and consumer surplus displayed in Figure \ref{sw_W} 
correspond to open access.
Here revenue decreases, but for $\beta=0$ the revenue initially
increases slightly as $W$ increases from zero (not shown).

We can obtain further insight by letting $W$ become large.
%Next we consider the scenario in which 
%outcome with $N$ symmetric SPs when 
%where the amount of new intermittent bandwidth $W$ becomes large. 
In this limit, some of the expressions simplify, easing the analysis.
For a given number of SPs $N$, taking this limit and using the equilibrium expressions
in the appendices, the total mass of customers served is given by 
\[
\rho(N) = \bar{x}(N) + \bar{w}(N) = \frac{B}{B+B/N +2(1-\alpha)}.
\]
This does not depend on $\beta$ and so if there 
is sufficient shared bandwidth, it does not matter how it is allocated.
The social welfare for licensed and open access shared bandwidth therefore
become the same as illustrated in Fig. \ref{sw_W}.
This is intuitive since as $W\rightarrow \infty$ the congestion externality 
disappears in the shared spectrum, so open access and licensed access
provide the same value to consumers.
Note that $\rho(N)$, and hence consumer welfare, 
increases with $N$ and approaches the asymptote
\[
\bar{\rho} =\frac{B}{B+2(1-\alpha)}.
\]

In the limit of large $W$, the social welfare as a function of $N$ is given by
\begin{align*}
SW(N)  &= \rho(N) -\frac{\rho^2(N)}{2}\left(1 + \frac{2( 1-\alpha)}{B}\right).
\end{align*}
This expression is an increasing function of $\rho(N)$ for $\rho(N) \leq \bar{\rho}$. Hence, $SW(N)$ is also increasing with $N$ and approaches the limiting value
\[
\frac{B}{2(B + 2(1-\alpha))} = \frac{\bar{\rho}}{2}.
\]
Note that $\bar{\rho}$ is a strictly increasing function of $\alpha \in [0,1]$ 
and for $\alpha =1$, we have $\bar{\rho} = 1$, meaning the entire market is served,  
resulting in a social welfare of $\frac{1}{2}$, 
which is the maximum possible for the assumed inverse demand.
For $\alpha <1$, we have $\bar{\rho} <1$, meaning that even 
with an unbounded amount of shared spectrum, 
some users are not served due to the intermittent nature 
of that spectrum, so that some potential welfare is not obtained.

As previously noted for arbitrary $W$, 
the previous results show that when $\alpha <1$ and $N\rightarrow \infty$, 
the aggregate profit of the SPs is strictly positive. 
In this case, the limiting aggregate firm profit is given by
\[
\bar{\rho}  - \bar{\rho}^2\left(1 + \frac{( 1-\alpha)}{B}\right) =
\frac{B(1-\alpha)}{(B+2(1-\alpha))^2}.
\]
Differentiating this with respect to $\alpha$, it can be seen that for $B <2$, the aggregate firm profit first increases with $\alpha$ and then decreases, with the maximum firm profits occurring when $\alpha = 1-B/2$. For $B\geq 2$, aggregate firm profits decrease with $\alpha$, and so the maximum occurs when $\alpha = 0$. In other words, given any value of $B$, the SPs would prefer that the shared spectrum is intermittent, and if $B$ is large enough, they would prefer that the shared spectrum is never available.  Adding new spectrum to the market reduces congestion, but also intensifies competition.
The latter effect becomes more pronounced the less intermittent the spectrum becomes 
and apparently dominates the impact on the providers' profits.

\section{Asymmetric Providers}\label{asymSP}
%NOTE: This should be combined with Vijay's model for one big provider
%and many small providers.**\\
To provide insight into the effects of asymmetric (large and small) SPs
with different amounts of bandwidth, we now consider the following
two scenarios:
\begin{itemize}
\item There is a single SP with proprietary bandwidth $B_1$.
A second band $B_2$ is split evenly among $N$ small SPs, 
where $N$ is assumed to be large.
\item
Bands $B_1$ and $B_2$ are each split among $N$ SPs.
We will assume that $B_1 \geq B_2$.
\end{itemize}
Varying $B_1$ relative to $B_2$ then captures varying degrees
of asymmetry. The two scenarios differ in the amount of competition
experienced by the SP(s) with the larger bandwidth allocation.

As before, the shared band is split between 
an open access part (bandwidth $\beta W$)
and proprietary part (bandwidth $(1-\beta) W$). 
The open access part is shared among all SPs, large and small.
The proprietary part is further split into two sub-bands 
with bandwidths $W_1$ and $W_2$ allocated to the large and
small SPs, respectively.
The shared band is intermittently available, so that
a small SP has proprietary bandwidth $B_2/N$, which is always available,
plus $W_2/N$, which is available with probability $\alpha$.

For the first scenario, let $x_1 (N)$ and $w_1 (N)$ denote the quantities served
by the large SP in its proprietary and open access spectrum, respectively.
The corresponding quantities for the $i^{\mathrm{th}}$ small SP are 
$x_{2,i} (N)$ and $w_{2,i} (N)$, respectively. By symmetry $x_{2,i} (N)$
and $w_{2,i} (N)$ are independent of $i$. 
As $N\to\infty$, the equilibrium quantities are defined as
\begin{equation}
(x_1^*,w_1^*,x_2^*,w_2^*) = \lim_{N\to\infty} [x_1(N),w_1(N),N x_{2,i}(N), N w_{2,i}(N) ]
\end{equation}
It is shown in Appendix \ref{app:asym} that those quantities
are the solution to a set of four linear equations.
Similarly, in the second scenario, $x_1(N)$ and $w_1(N)$
are replaced by $x_{1,j}(N)$ and $w_{1,j}(N)$, where $j$ denotes
an SP in the larger group. From symmetry those quantities 
are independent of $j$, and as $N\to\infty$,
the corresponding equilibrium quantities are
\begin{equation}
(x_1^*,w_1^*,x_2^*,w_2^*) = \lim_{N\to\infty} 
[N x_{1,j} (N),N w_{1,j} (N),N x_{2,i}(N), N w_{2,i}(N) ]
\end{equation}

As for the scenario with two asymmetric SPs, here also a larger SP
does not always use the open access spectrum. In addition, for the second asymmetric scenario
considered here with many larger SPs, a {\em smaller} SP may not use
the open access spectrum. The associated conditions are stated next.
\begin{theorem}
\label{prop:asym}
For the first scenario with asymmetric SPs, if
\begin{align}\label{asympcond1}
B_1 + W_1 + 2 (1-\alpha) \frac{W_1}{B_1} > 
%2 (1-\alpha) \frac{W_2}{B_2} + 4 (1-\alpha) \frac{W_3}{B_2},
%2 (1-\alpha) \frac{W_2}{B_2} + 4 (1-\alpha) \frac{\beta W}{B_2},
2 (1-\alpha) \frac{2\beta W + W_2}{B_2}
\end{align}
then there exists an $N^*$ such that for all $N \geq N^*$, SP 1 does not use
the open access spectrum, {\em i.e.}, $w_1^*(N)=0$. 

For the second scenario, a larger (smaller) SP $i$ does not use the open access spectrum 
for all $N\geq N^{**}$ (for some $N^{**}$) when
\begin{equation}
\frac{W_i}{B_i} > \frac{2 \beta W + W_j}{B_j},~~~i \neq j.
\label{asympcond2}
\end{equation}
where $j$ corresponds to a smaller (larger) SP.
\end{theorem}
The proof can be found in Appendix \ref{structure}. 

%is analogous to the proof of Proposition \ref{prop2} and Theorem \ref{thm:asym}.

The first condition \eqref{asympcond1} resembles, but is not identical to the condition 
in Theorem \ref{thm:asym}.
This is because here a portion of the licensed spectrum is also intermittent.
Note that for $\alpha=1$ the larger SP {\em always} vacates the open access spectrum. 
Also, for $\alpha=1$, the price in the shared spectrum is zero,
which is also true for the analogous model with Bertrand competition \cite{nguyen2014cost}.
The first condition in Theorem \ref{prop:asym} is satisfied when $\beta$ 
and $W_2$ are small. In that case, the smaller SPs congest
the open access band, lowering the price, and thereby make it less desirable 
for the large SP(s). For small $\beta$ the second condition \eqref{asympcond2} becomes
$W_i/B_i > W_j/B_j$. If $W_i > W_j$, then the condition can be satisfied with
$B_i < B_j$, i.e., the smaller SPs vacate the open access band.
This is due to competition among the larger SPs, which causes them 
to shift traffic to the open access band,
increasing congestion in that band and lowering the price
so that the smaller SPs have no incentive to use it.

The condition \eqref{asympcond2}
does not depend on $\alpha$, in contrast to \eqref{asympcond1},
because for large $N$, the additional congestion 
caused by intermittency is bounded, and is shared
among the $N$ large SPs. Hence that additional congestion 
does not significantly affect an individual SP. 
As $\alpha$ decreases, the threshold $N^{**}$ must increase
in order for the condition to apply.

%Figure \ref{asym_beta_b} shows that the total welfare may not be
%monotonic or convex for intermediate values of $n_j$.
%Here the maximum occurs for $\beta$ between zero and one indicating that
%only part of the band should be shared. This type of example
%arises only with asymmetric bandwidth allocations. Additional 
%experimentation with model parameters indicate that examples
%where partial sharing is optimal are not straightforward to find.
%Furthermore, as indicated in this example, the benefit from
%partial sharing versus full sharing is relatively small.
%\begin{figure}
%\centering
%\begin{subfigure}{3.5in}
%  \centering
%  \includegraphics[width=3.2in]{Plots/sw_beta_nj1_B19_crop.pdf}
%  \caption{$n_j=1$ and $n_j=3000$}
%  \label{asym_beta_a}
%\end{subfigure}%
%\begin{subfigure}{3.5in}
%  \centering
%  \includegraphics[width=3.2in]{Plots/sw_beta_nj30_B19_crop.pdf}
%  \caption{$n_j=30$}
%  \label{asym_beta_b}
%\end{subfigure}
%\caption{Revenue, consumer surplus, and total welfare versus
%amount of shared bandwidth $\beta$ with different numbers of SPs $n_j$.}
%\label{asym_beta}
%\end{figure}
%
%{\color{red}  $B1$ should be $B_1$.}

\begin{figure}
%\centering
\begin{subfigure}{3.7in}
%  \centering
  \includegraphics[width=3.7in,height=2.5in]{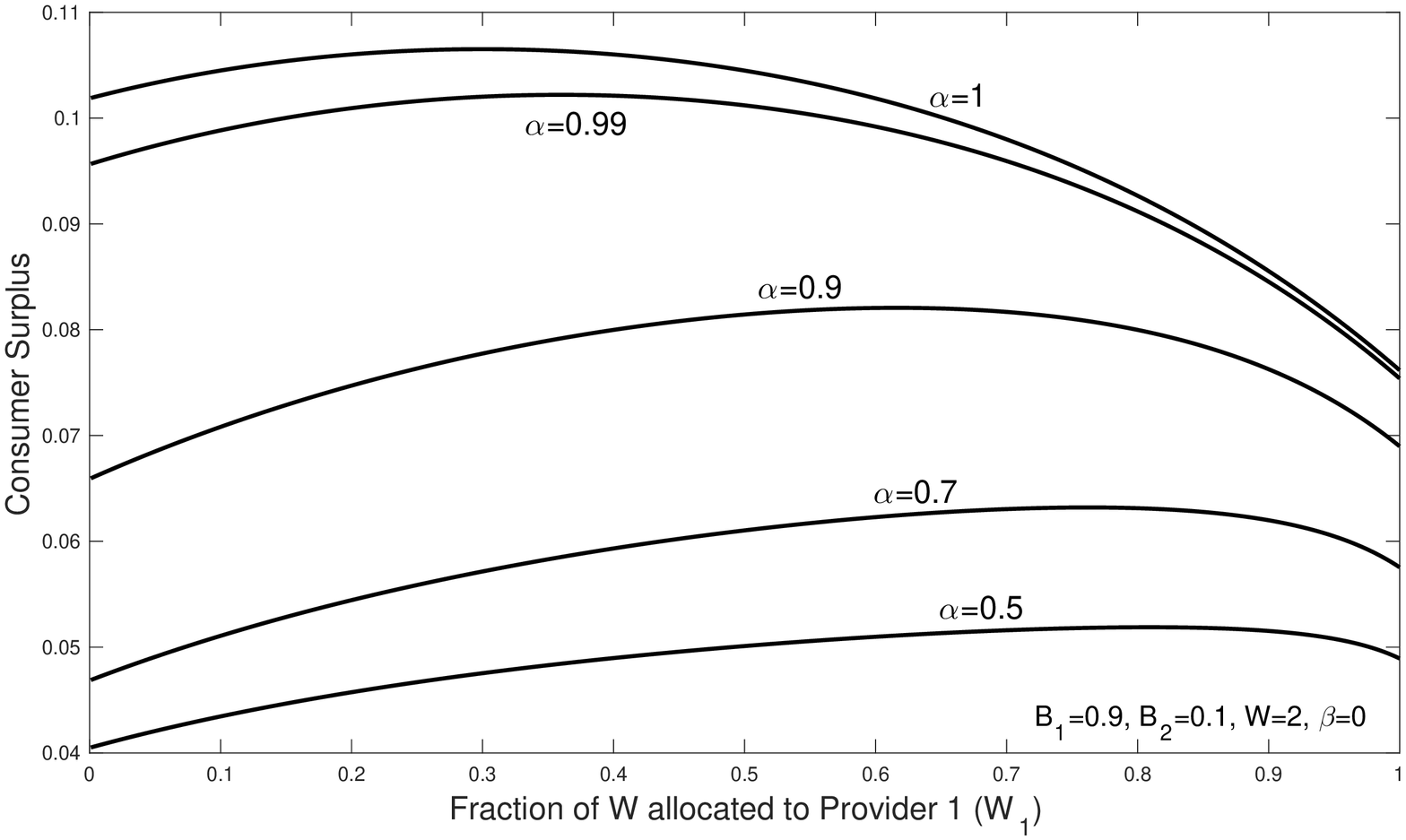}
  \caption{$N=2$}
  \label{asym_W1_a}
\end{subfigure}%
\begin{subfigure}{3.7in}
%  \centering
  \includegraphics[width=3.7in,height=2.5in]{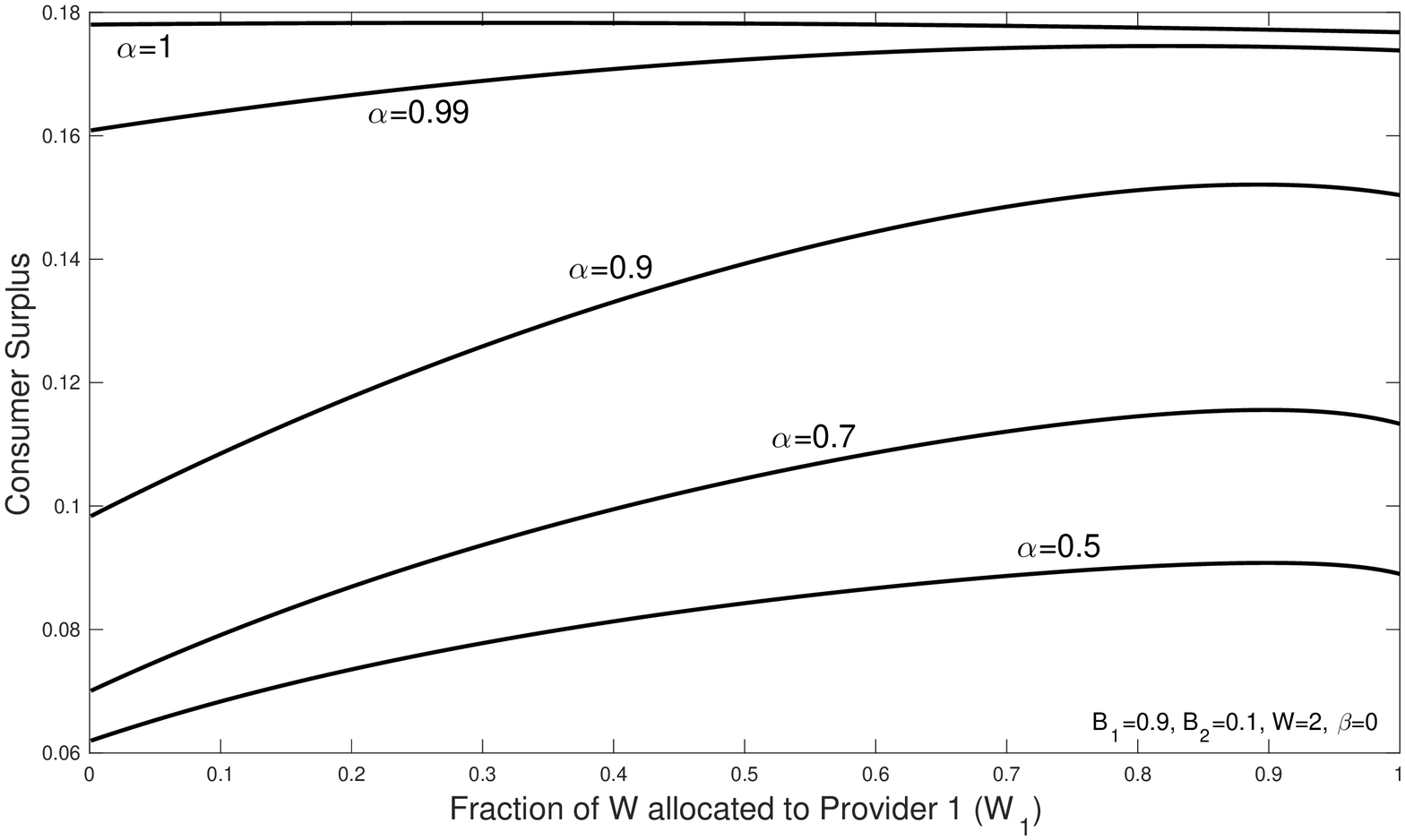}
  \caption{$N=60$}
  \label{asym_W1_b}
\end{subfigure}
\caption{Consumer surplus versus the fraction of shared bandwidth assigned
to SP 1 ($W_1/W$) with different values of $\alpha$.}
\label{asym_W1}
\end{figure}

Fig. \ref{asym_W1} illustrates how the split of the shared band $W$
into $W_1$ and $W_2$ affects consumer surplus. Here $B_1=0.9$, $B_2=0.1$,
$W=2$, $\beta=0$ (all of $W$ is split between the SPs),
and plots are shown for different values of $\alpha$.
The figures show consumer welfare as a function of $W_1/W$
for $N=2$ (Fig. \ref{asym_W1_a}) and $N=60$ (Fig. \ref{asym_W1_b}).
As $\alpha$ increases, Fig. \ref{asym_W1_a} shows that
the fraction of bandwidth that maximizes consumer surplus
shifts to the left.
This is due to the tradeoff between the larger SP's
ability to handle intermittent traffic, and competition.
That is, when $\alpha$ is small,
most of the shared band should be allocated to the larger SP,
since the larger SP is better able to handle the intermittent
availability of the shared band. As $\alpha$ increases, so that the band
becomes more reliable, the consumer surplus increases by shifting
bandwidth to the smaller SP to increase competition.
In contrast, Fig. \ref{asym_W1}b shows that with many competing
SPs it is always best to give most of the shared bandwidth
to the larger SPs, independent of $\alpha$.

%{\color{red}  $B1$ should be $B_1$.}

As for the symmetric case, numerical examples show that
%Figure \ref{asym_beta} shows plots of total provider revenue,
%consumer surplus, and total welfare versus the amount of shared
%bandwidth $\beta$ with different numbers of SPs $n_j$. 
%{\color{red}  what is $n_j$?}
%For this example $B_1=0.9$, $B_2=0.1$, and $\alpha=0.8$.
%Figure \ref{asym_beta_a} shows 
social welfare decreases with $\beta$ when $N \to \infty$, 
which is the same as for the symmetric case. 
%(For this example, the curve has converged for $n_j>3000$.) 
In contrast, for $N=2$ the total welfare {\em increases}
with $\beta$, as discussed in Section \ref{sec:TwoSPs}.
%(This will be shown in Section \ref{twoproviders}.) 
Hence at these extreme values designating the entire band $W$
as open access ($N=2$) or proprietary ($N \to \infty$) maximizes total welfare.
Additional numerical examples indicate that this is also
true for arbitrary $N$.

\section{Extensions to General Latency and Demand}
\label{sec:marginal}
\Xomit{
{\color{blue} Given that our model allows for parts of the intermittent secondary band to be licensed, when the secondary band is available, the total licensed bandwidth of a SP, and hence, the latency cost experienced by customers served on the licensed band could change. Hence, we will assume that when the intermittent band is available, the latency function is given by $\ell_{i,w}(\cdot)$, and $\ell_i(\cdot)$ when not available.} {\color{red} Existence, uniqueness and potential function results modified with this change. This is needed for the arbitrary $\beta\in [0,1]$ partitions that we're discussing at present.}
}

In this section we establish existence of a unique equilibrium 
for the Cournot game with more general demand and latency functions. 
The model allows the shared band
to be split between licensed and open access. 
When the shared band is available, the total licensed bandwidth of a SP changes
and hence so does the latency cost experienced by customers served on the licensed band.
%\footnote{ Recall Section \ref{symprov} and the parameter $\beta\in [0,1]$.} 
In addition, with a linear inverse demand function, linear latencies in the open access bands and general convex increasing latencies in the licensed bands, we also prove that the game
is a potential game.\footnote{While the result in \cite{perakis2014efficiency} proves the existence of a unique equilibrium for linear inverse demand and latencies (with a more general shared model), it does not provide the potential game characterization.}
This fact is used in deriving some of our earlier results.

Assume that when the intermittent band is available, 
the latency function is given by $\ell_{i,w} (\cdot)$, 
and $\ell_i (\cdot)$ when not available. 
%Existence, uniqueness and potential function results modified with this change. 
%This is needed for the arbitrary β ∈ [0,1] partitions that we’re discussing at present.
\begin{theorem}
\label{thm:existence}
For the Cournot game with $N \geq 2$ providers, each with proprietary spectrum
and additional intermittently available shared spectrum, if the inverse demand
$P(\cdot)$ is concave decreasing, and the latencies $\ell_i (\cdot)$, $\ell_{i,w}(\cdot)$ and $\ell_w (\cdot)$
are convex increasing, then an equilibrium always exists. 
The equilibrium is unique if either $P'(0) < 0$ and $\ell_w' (0) > 0$
or $\ell_i' (0) > 0$ and $\ell_{i,w}'(0)>0$ for all $i=1,2, \dotsc, N$.
\end{theorem}

In the absence of open access spectrum, Theorem \ref{thm:existence} holds without the
condition on $\ell_w' (\cdot)$ and $\ell_{i,w}'(\cdot)$ for all $i=1,2,\dotsc,N$. 
\Xomit{ The proof is given in Appendix \ref{sec:potential}
and consists of showing that this is a concave game and 
applying results from Rosen \cite{Rosen}.
For the special case where the inverse demand is linear decreasing,
{\em i.e.}, $P(y) = 1 - \gamma y$, the latency in the shared spectrum is linear,
and $\ell_w (w) = w/W$, it can be verified that the game 
is a potential game with concave potential
\begin{align*}
\label{eq:potfn}
\Phi(\mathbf{y},\mathbf{w})& = \sum_{i=1}^N y_i - \gamma \sum_{i=1}^N y_i^2 - \gamma \sum_{i=2}^N \sum_{j=1}^{i-1} y_i y_j  - (1-\alpha) \sum_{i=1}^N l_i(y_i) y_i - \alpha \sum_{i=1}^N l_{i,w}(y_i-w_i) (y_i-w_i) \\
& \quad 
- \frac{\alpha}{W} \sum_{i=1}^N w_i^2 - \frac{\alpha}{W} \sum_{i=2}^N\sum_{j=1}^{i-1} w_i w_j
\end{align*}
where $y_i = x_i + w_i$ is the total quantity served by provider $i$.
With $N=2$ providers,  
the game is supermodular {\color{red} for ....} See  Appendix \ref{sec:potential} for proofs.
Hence, for concave decreasing inverse demand and increasing convex latency, 
there exists a unique (generalized) equilibrium, which can be computed via a sequence
of best-response updates.

For linear latencies and $W=0$ this simplifies to
\begin{align*}
\Phi(\mathbf{x})=\sum_{i=1}^N x_i - \sum_{i=1}^N x_i^2 -\sum_{i=1}^N x_i l_i(x_i) - \sum_{i=2}^{N}\sum_{j=1}^{i-1} x_i x_j .
\end{align*}
}
%giving a relatively simple way to compute the equilibrium.
When bandwidths $W_1, W_2, \dotsc, W_N$ 
with intermittent availability $\alpha$ are added,
this is equivalent to the set of {\em non}-intermittent 
bands 
%$\tilde{B}_1,\tilde{B}_2,\dotsc,\tilde{B}_N$ 
$T_1,\cdots,T_N$, where $T_i$ is give in \eqref{Ti}.
%with $\tilde{B}_i = B_i \tfrac{B_i +W_i}{B_i+(1-\alpha) W_i}$.
%\textbf{Most likely the supermodular game part for $N=2$ holds too.}
With linear decreasing inverse demand, the existence
and uniqueness of an equilibrium
follows from Proposition 2 in \cite{perakis2014efficiency}. 
%(The formulation as a potential game does not appear there.)
%The extension to general $P(\cdot)$
%is used in Section \ref{sec:marginal}.

We now assume two SPs with propriety spectrum {\em only}, 
{\em i.e.}, any shared spectrum is licensed and always available,
so that we can assume $W=0$.
%there is no additional shared spectrum ($W=0$).
The proofs of the following propositions are in Appendix \ref{app:marginal}.

\begin{prop} 
\label{marg:prop1}
Given an equilibrium (interior point) with two providers, concave
decreasing inverse demand, and convex increasing latency, 
if a marginal amount of bandwidth is given to provider $k$, then
a sequence of best responses converges to a new equilibrium
in which the quantity $x_k$ and revenue $R_k$ each increase
and $x_{-k}$ and $R_{-k}$ each decrease.
\end{prop}

According to Theorem \ref{thm:existence}, the sequence of best
responses must converge to a unique equilibrium. This extends
Theorem 3.2, and states that in this more general setting 
an increase in one provider's bandwidth again causes a decrease
in the competitor's quantity and revenue.

\begin{prop}
\label{marg:prop2}
For the scenario in Prop. \ref{marg:prop1}, giving a marginal amount
amount of bandwidth to SP $k$ increases both consumer surplus and
total welfare.
\end{prop}

This states that although from Proposition \ref{marg:prop1}, 
adding this marginal bandwidth increases $x_k$ and decreases $x_{-k}$,
the {\em total} quantity of customers served increases.
Similarly, although the revenue $R_{-k}$ decreases, the total welfare increases.

Suppose now that we wish to give the bandwidth to the SP which will 
increase consumer surplus the most. That means allocating the bandwidth 
to maximize the total incremental quantity customers served.
In general, this depends on the derivative $P'(\cdot)$ and
second derivative $\ell '' (\cdot)$, and is somewhat complicated
(see Appendix \ref{app:marginal}); however, for linear
latencies $\ell_k (x) = c_k x$, it reduces to finding
\begin{equation}
\label{eq:marg1}
\arg \max_k - \frac{c_k x_k}{B_k^2} \left( P' (x_1 + x_2 ) 
- \frac{2 c_{-k}}{B_{-k}} \right) 
\end{equation}
Further constraining $P(x) = 1 - ax$, and using the best response
conditions for $x_k$, $x_{-k}$, the bandwidth should be given
to agent $k$ if
\begin{equation}
\label{eq:marg2}
B_{-k} > \sqrt{\frac{c_{-k}}{c_k}} B_k + \frac2a ( \sqrt{c_k c_{-k}} - c_{-k} )
\end{equation}
Otherwise, it should be given to agent $-k$. 
%The result is similar
%Theorem \ref{thm:asym}, where the `smaller' SPs only make use of the additional bandwidth. In Theorem \ref{thm:asym} this was an equilibrium outcome, 
Recall that when allocating additional {\em intermittent} spectrum ($\alpha<1$),
the consumer surplus is given by \eqref{cs}-\eqref{x1+x2}.
Here we effectively have $\alpha=0$ so that $T_i=B_i+W_i$, which replaces $B_i$
in the preceding condition. When $c_k=c_{-k}$ the condition reduces
to $B_{-k} > B_k$, so that any marginal bandwidth should attempt 
to equalize the bandwidth allocation.
If $c_k < c_{-k}$, however, the allocation is biased towards SP $k$,
which provides lower latency.
%the argument above shows that it also improves consumer surplus. Furthermore, the notion of  `big' and `small' depend not just on bandwidth but properties of the latency as well.

\section{Conclusions}\label{conc}
We have presented a model for sharing intermittently available spectrum that captures
licensed and open access sharing modes, congestion as a function of offered load,
and competitive pricing for spectrum access.
Our analysis suggests that allocating shared bandwidth as open access 
is better for consumer surplus than licensing the bandwidth for exclusive use.
While latencies will be high, that is offset by lower prices, which
has the effect of expanding the demand for services.
Allocating additional bandwidth as licensed is good for revenue, 
because SPs generally choose to lower congestion by raising prices.
%These higher prices also limit the demand for services. 
The trade-off among revenue, consumer surplus, and congestion depends greatly 
on the market structure. With many SPs, competition may be enough
so that total welfare (revenue plus consumer surplus) is maximized
by licensing the intermittent bandwidth. With asymmetric SPs having
different amounts of bandwidth, it is also possible that only a subset of the SPs
use the open access band to maintain higher prices, thereby containing congestion.

The model might be enhanced in several different ways.
We have not directly accounted for investment, which may be used to mitigate
congestion, although we have shown that our main conclusions are robust 
with respect to a congestion penalty for open access.
We have also generally assumed that access to the shared band is free,
and have not considered pricing mechanisms, which could
be used to allocate the shared spectrum as a combination of licensed and open access.
Those features might also be combined with an extended model that allows SPs 
without proprietary spectrum to bid for open access spectrum, potentially combining
both price and quantity competition.

%{\thanhedit
%Deciding on allocating intermittent spectrum is tricky. Two methods are largely discussed: allocating as licensed (by auction);  as unlicensed; or mixture of the two. 

%We show that this problem is tricky. It depends not only on the level of intermittency, but also on many things. But general, we learn that:
%\begin{itemize}
%\item using auction can have reverse effect on welfare.
%\item because of congestion effect,   prices is lower in unlicensed bands, congestion is higher. Thus allocating as unlicensed is good for  market expansion with the cost of quality.
%\item allocating as licensed is good for revenue, because SPs  can manage congestion better than unlicensed.
%\item the trade-off between these two forces is not trivial. It depends greatly on the existing market structure: many small SPs, or few large SPs and many small ones?
  %\end{itemize}
%}

\bibliographystyle{plain}
\bibliography{ref}

%\section{Licensed Secondary Regime }\label{status}
%\input{nowhitespace}

%\section{Unlicensed Secondary Regime}
%\input{specialcase}
%\subsection{The case $K_i = B$}
%

%\subsection{Two Service Providers}
%\input{whitespace}

%\section{more than 2 SPS: Supermodularity and potential properties}
%\input{general}

\section{Appendix}
%We collect together the proofs of the results claimed in the body of the paper.
%\input{appendix}
\subsection{Bidding for shared spectrum: licensed versus open access}
\label{app:bids}
We consider the scenario in which two SPs can bid on the shared bandwidth $W$,
and show that they may prefer that $W$ be allocated as open access rather than licensed. 
Assume that $B_1=B_2$, and that the incumbent of the shared band wishes
to distribute $W=1$. It can offer it as open access or allocate 
it entirely to one of the providers.
The table below records the revenue each provider makes under each possibility.
\\
\\
\begin{center}
\begin{tabular} {|l|c|c|c|c|} \hline
$\alpha$ & pre-allocation & large & small & open access \\ \hline
0.1 & 0.08 &  0.096 & 0.052 & 0.063  \\ \hline
0.5& 0.08 & 0.102 & 0.057 & 0.068 \\ \hline
0.9 & 0.08 & 0.11 & 0.064 & 0.075 \\ \hline
\end{tabular}
\end{center}
The column labeled `pre-allocation' lists the revenue of each SP
before the allocation of additional bandwidth. The column labeled `large' records 
the revenue of the SP that receives the entire one unit of additional bandwidth. 
The column labeled `small' is the revenue of the SP who did not receive the additional bandwidth. 
The last column is the revenue of each SP when the additional unit of bandwidth 
is offered as open access. Each row corresponds to various levels of $\alpha$.

Consider, for example, the case $\alpha = 0.5$. 
Suppose the one unit of additional bandwidth is allocated in its entirety via an ascending auction.
If the current price is $p$, an SP should remain active provided:
$$0.102 - p \geq 0.057\,\, \Rightarrow \,\, p \leq 0.045.$$
Hence, each SP should remain active until the price reaches 0.045. 
The winner will earn a profit of 0.102 - 0.045 = 0.057. 
Notice, the profit of each SP is higher in the open access regime. 
Hence, when paying for the additional bandwidth
each SP would prefer open access sharing over having 
the additional bandwidth as licensed to themselves.

\subsection{Proof of Theorem \ref{thm3.1}}
%{\color{red} Ricky, Thanh: The proofs of Thms 3.2 and 3.3 appear to be incomplete.}
Theorem \ref{thm3.1} follows in the standard way by deriving the reaction functions of each provider and determining their intersection. Hence, many of the details are omitted. The revenue of provider $i$, denoted $R_i$ is:
$$p_i\left(1 - x_i\left(1+\frac{1}{T_i} \right) - x_{-i}\right).$$
Compute
 $\partial R_i / \partial x_i$ for each $i$ and set to zero.  The solution of this pair of first order conditions is unique. The corresponding equilibrium quantities and prices for provider 1 are
%Recall the non-intermittent bandwidth case from section \ref{twolic}. 
%The equilibrium quantities are given by:
\begin{align*}
x_1^*=\frac{T_1}{T_1 \frac{3T_2+4}{T_2+2}+\frac{4T_2+4}{T_2+2}}, \quad p_1^*=\frac{T_1+1}{T_1 \frac{3T_2+4}{T_2+2}+\frac{4T_2+4}{T_2+2}}
\end{align*}
so that $x_1^*=\tfrac{T_1}{b T_1 + a}$ and $p_1^*=\tfrac{T_1+1}{b T_1 +a}$ where
\begin{align}\label{eq:abval}
b=\frac{3T_2+4}{T_2+2}=3-\frac{2}{T_2+2} \in [2,3), \quad a=\frac{4T_2+4}{T_2+2}=4-\frac{4}{T_2+2} \in [2,4)
\end{align}
\subsection{Proof of Theorem \ref{status-rev}}
To prove Theorem \ref{status-rev}, recall that the revenue of provider $1$ is given by $R_1^*=p_1^* x_1^*=\tfrac{T_1(T_1+1)}{(bT_1+a)^2}$ so that
\begin{align*}
\frac{\partial R_1}{\partial T_1} & = \frac{(2T_1+1)(bT_1+a) - 2 b T_1 (T_1+1)}{(b T_1 +a)^3}=\frac{(2a-b)T_1+a}{(bT_1+a)^3} \\
\frac{\partial^2 R_1}{\partial T_1^2} & = \frac{(2a-b)(bT_1+a)-3b\big((2a-b)T_1+a\big)}{(bT_1+a)^4}=\frac{-2b (2a-b) T_1 - 2a (2b-a) }{(bT_1+a)^4},
\end{align*}
where $a$ and $b$ are defined in \eqref{eq:abval}. Using the expressions for $a$ and $b$, we get
\begin{align*}
2a-b=5-\frac{6}{T_2+2} \in [2,5), \quad 2b-a=2.
\end{align*}
Therefore, the revenue of provider $1$ is strictly concave and increasing in $T_1$ for any given value of $T_2$.

Both $b$ and $a$ are increasing in $T_2$ and so it follows that $R_1^*$ is decreasing in $T_2$.

\subsection{Proof of Theorem \ref{thm3.4}}

%{\color{red} Ricky, Vijay: This needs to be sorted out and edited.}
If an interior equilibrium exists, then:
\begin{align*}
p_i & = 1 - x_i (1+\tfrac{1}{B_i}) - x_{-i} - w_i (1+\tfrac{(1-\alpha)\beta}{B_i}) - w_{-i} \\
p_i^w & = 1- x_i (1+\tfrac{1-\alpha}{B_i}) - x_{-i} - w_i (1+\tfrac{\alpha}{W}+\tfrac{(1-\alpha)\beta}{B_i}) - w_{-i} (1+ \tfrac{\alpha}{W})
\end{align*}
Using this, the revenue of provider $i$ is:
\begin{align*}
& R_i = p_i x_i + p_i^w w_i \\
& = (1-x_{-i}-w_{-i}) x_i - x_i w_i (2+\tfrac{(1-\alpha)(1+\beta)}{B_i}) - x_i^2 (1+\tfrac{1}{B_i}) + (1-x_{-i}-w_{-i}(1+\tfrac{\alpha}{W})) w_i \\
& \qquad - w_i^2 (1+\tfrac{\alpha}{W}+\tfrac{(1-\alpha)\beta}{B_i})
\end{align*}
Assuming the revenue is jointly concave in $(x_i,w_i)$ (which is true for $\beta=1$ the case of interest), the best response functions are obtained by setting the following (partial) derivatives to $0$, namely,
\begin{align*}
\frac{\partial R_i}{\partial x_i} & = 1-x_{-i}-w_{-i} - w_i (2+\tfrac{(1-\alpha)(1+\beta)}{B_i}) - 2 x_i (1+\tfrac{1}{B_i}) \\ 
\frac{\partial R_i}{\partial w_i} & = 1-x_{-i}-w_{-i}(1+\tfrac{\alpha}{W}) - x_i (2+\tfrac{(1-\alpha)(1+\beta)}{B_i}) - 2 w_i (1+\tfrac{\alpha}{W}+\tfrac{(1-\alpha)\beta}{B_i}).
\end{align*}

In the symmetric case of $B_1=B_2=B/2$, we search for a symmetric equilibrium using the above to get the following linear equations
\begin{align*}
x (3+\tfrac{4}{B}) + w (3+\tfrac{2(1-\alpha)(1+\beta)}{B}) & =1 \\
x (3+\tfrac{2(1-\alpha)(1+\beta)}{B}) + w (3+\tfrac{3\alpha}{W}+\tfrac{4 (1-\alpha)\beta}{B}) & =1 
\end{align*}

Solving this yields the quantities in the theorem.
%following unique solution (solution exists for $\beta=1$) for $\beta=1$,
%\begin{align*}
%x & = \frac{\frac{3}{W}}{\frac{9}{W}+\frac{6}{K}+\frac{6}{K W}+\frac{4 (1-\alpha)}{K^2}} \\
%w & = \frac{\frac{2}{K}}{\frac{9}{W}+\frac{6}{K}+\frac{6}{K W}+\frac{4 (1-\alpha)}{K^2}}
%\end{align*}
The resulting prices are
\begin{align*}
p_1=p_2 & = 1- x (2+\tfrac{2}{B}) - w (2+\tfrac{2(1-\alpha)}{B}) \\
p_1^w=p_2^w & = 1- x (2+\tfrac{2(1-\alpha)}{B}) - w (2+\tfrac{2\alpha}{W}+\tfrac{2(1-\alpha)}{B})
\end{align*}
From the equations of the equilibrium, it can be gleaned that the prices are positive. 
Thus, an interior equilibrium exists.

\subsection{Proof of Theorem \ref{corr:01}.}
\label{app:manyusers}
For the analysis we assume that $\min\big(B+(1-\beta)W, \beta W\big)>0$. The results for the equilibrium quantities when this condition does not hold follow by continuity with the additional assumption that $B=0$ is necessarily accompanied with $\alpha=1$.

The revenue of SP $i$ is
\begin{align*}
R_i =& p_i x_i + p_i^w w_i \\
 = &\left (1-\sum_{j=1}^N x_{j}-\sum_{j=1}^N w_{j}-\alpha N \frac{x_i}{B+(1-\beta)W} -(1-\alpha) N \frac{w_i+x_i}{B}\right)x_i\\
+&\left (1-\sum_{j=1}^N x_{j}-\sum_{j=1}^N w_{j}-\alpha \frac{\sum_{j=1}^N w_j}{\beta W}-(1-\alpha) N\frac{w_i+x_i}{B}\right)w_i
\end{align*}

Taking the derivative of $R_i$ with respect to $x_i$
\begin{align*}
\frac{\partial R_i}{\partial x_i}  = & \left(1-\sum_{j=1}^N x_{j}-\sum_{j=1}^N w_{j}-\alpha N \frac{x_i}{B+(1-\beta) W} -(1-\alpha) N \frac{w_i+x_i}{B}\right) \\
& \quad -\left (1+\alpha N \frac{1}{B+(1-\beta) W}+(1-\alpha) N \frac{1}{B}\right ) x_i-(1+  \frac{(1-\alpha) N}{B})w_i
\end{align*}
We will derive a symmetric equilibrium. It will follow from a subsequent theorem, Theorem \ref{thm:existence}, that the unique equilibrium is symmetric. If we set $x_i = x$ and $w_i = w$ for all $i$, we get:
\begin{align*}
1  = & \left(N x+ N w +\alpha N \frac{x}{B+(1-\beta) W} +(1-\alpha) N \frac{x+w }{B}\right) \\
& \quad +\left (1+\alpha N \frac{1}{B+(1-\beta) W}+(1-\alpha) N \frac{1}{B}\right )x+(1+  \frac{(1-\alpha) N}{B})w
\end{align*}
This implies 
\begin{equation}\label{eq:dev1}
1  = x\left(1+N+ 2N \frac{\alpha}{B+(1-\beta) W} +2N \frac{(1-\alpha)}{B}\right) +w\left (1+N+2N\frac{1-\alpha}{B} \right )
\end{equation}

Taking derivative with respect to $w_i$
\begin{align*}
\frac{\partial R_i}{\partial w_i}  = &- \left(1+\frac{(1-\alpha)N}{B}\right)x_i\\
+&\left (1-\sum_{j=1}^N x_{j}-\sum_{j=1}^N w_{j}-\alpha  \frac{\sum_{j=1}^N w_j}{\beta W}-(1-\alpha) N \frac{w_i+x_i}{B}\right)-\left(1+\frac{\alpha}{\beta W} + \frac{(1-\alpha)N}{B}\right)w_i
\end{align*}

Using an argument similar to that above we get 
\begin{equation}\label{eq:dev2}
1  =  x\left(1+N+2N\frac{1-\alpha}{B}\right)+w\left( 1+\frac{\alpha}{\beta W}+ N+ N\frac{\alpha}{\beta W}+2N\frac{1-\alpha}{B}\right)
\end{equation}
Solving for the equilibrium using \eqref{eq:dev1} and \eqref{eq:dev2}, we obtain
\begin{align*}
\bar{x}(N) & = \frac{\frac{\alpha (N+1)}{\beta W}}{\Big(1+N +\frac{2(1-\alpha) N}{B} \Big)\Big(\frac{2\alpha N}{B+(1-\beta) W} + \frac{\alpha (N+1)}{\beta W}\Big) + \frac{2\alpha N}{B+(1-\beta) W}\frac{\alpha (N+1)}{\beta W}},\\
\bar{w}(N) & = \frac{\frac{2\alpha N}{B+(1-\beta) W}}{\Big(1+N +\frac{2(1-\alpha) N}{B} \Big)\Big(\frac{2\alpha N}{B+(1-\beta) W} + \frac{\alpha (N+1)}{\beta W}\Big) + \frac{2\alpha N}{B+(1-\beta) W}\frac{\alpha (N+1)}{\beta W} }.
\end{align*}
Rearranging we get
\begin{align}
N\bar{x}(N) & = \frac{B^2+(1-\beta)BW}{\left(B\frac{1+N}{N}+2(1-\alpha)\right)\left(2\beta W \frac{N}{N+1}+B+(1-\beta)W\right)+2\alpha B}, \label{eq:barxn}\\
N\bar{w}(N) & = \frac{2\beta B W \frac{N}{N+1}}{\left(B\frac{1+N}{N}+2(1-\alpha)\right)\left(2\beta W \frac{N}{N+1}+B+(1-\beta)W\right)+2\alpha B}. \label{eq:barwn}
\end{align}
%
%Subtract from \eqref{eq:dev1}  \eqref{eq:dev2}, we obtain
%
%\begin{equation}\label{eq:xandw}
% \frac{nx}{B+W_1}=\frac{nw}{2W_2}.
%\end{equation}
The results then follow in a straight-forward manner by taking the limit as $N$ increases to infinity. We also note that the congestion in the shared band is exactly $\tfrac{2N}{N+1}$ times the congestion in the proprietary bands.
\QED
\subsubsection{Consumer Surplus}
\label{app:CS}
The total traffic carried $\rho(N)$ is then
\begin{align*}
\rho(N) & = N\bar{x}(N)+N\bar{w}(N) \\
& = \frac{B^2+\left(1+\beta \frac{N-1}{N+1}\right) BW}{\left(B\frac{1+N}{N}+2(1-\alpha)\right)\left(2\beta W \frac{N}{N+1}+B+(1-\beta)W\right)+2\alpha B} \\
& = \frac{B \left(B+ \left(1+\beta \frac{N-1}{N+1}\right) W\right)}{\left(B+ \frac{B}{N}+2(1-\alpha)\right)\left(B+ \left(1+\beta \frac{N-1}{N+1}\right) W\right)+2\alpha B},
\end{align*}
which is of the form $\frac{a f(\beta)}{(a+b)f(\beta)+c}$ where $f(\beta)$ is an increasing function of $\beta,$ and $a, b, c > 0$. Therefore, it is immediate that $\rho(N)$ is also an increasing function of $\beta$, and so is maximized at $\beta=1$.

Next we present the proof of Lemma \ref{4.4}, which studies consumer surplus with degraded shared spectrum. When we allow the shared spectrum to get degraded, i.e., reduce by a factor $d$, then we can use the formulae in \eqref{eq:barxn} and \eqref{eq:barwn} to analyze the equilibrium by setting the shared bandwidth $\beta W$ to $d \beta  W$. Define $d^*=\tfrac{2dN}{N+1}$. Then the congestion in the shared band is still $\tfrac{2N}{N+1}$ times the congestion in the proprietary band, i.e., denoting $(\bar{x}_\beta(N),\bar{w}_\beta(N))$ as the equilibrium quantities have the following expressions:
\begin{align*}
N\bar{x}_\beta(N) & = \frac{B^2+(1-\beta)BW}{\left(B\frac{1+N}{N}+2(1-\alpha)\right)\left(2d\beta W \frac{N}{N+1}+B+(1-\beta)W\right)+2\alpha B},\\
N\bar{w}_\beta(N) & = \frac{2d\beta B W \frac{N}{N+1}}{\left(B\frac{1+N}{N}+2(1-\alpha)\right)\left(2d\beta W \frac{N}{N+1}+B+(1-\beta)W\right)+2\alpha B}, \\
\Leftrightarrow \quad \frac{N\bar{w}_\beta(N)}{d\beta W} & =\frac{2N}{N+1} \frac{N\bar{x}_\beta(N)}{B+(1-\beta)W} ,
\end{align*}
where we used the increased latency in the shared band for the comparison. Using this we have the total traffic served $\rho_\beta(N)$ is then
\begin{align*}
\rho_\beta(N) & = N\bar{x}_\beta(N) + N\bar{w}_\beta(N) \\
& = \frac{B \left(B+ \left(1+\beta (d^*-1)\right) W\right)}{\left(B+ \frac{B}{N}+2(1-\alpha)\right)\left(B+ \left(1+\beta (d^*-1)\right) W\right)+2\alpha B},
\end{align*}
which is an increasing function of $\beta$ if $d^*>1$ and a decreasing function of $\beta$ if $d^*<1$. This conclusion then directly implies that the consumer surplus is maximized at $\beta=1$ if $d^*>1$ and at $\beta=0$ if $d^*<1$. 
\subsubsection{Total Surplus} 
Assuming no degradation of the shared band the revenue of SP $i$ is 
\begin{align*}
R_i 
 = &\left (1-N \bar{x}(N)- N \bar{w}(N)-\alpha N \frac{\bar{x}(N)}{B+(1-\beta)W} -(1-\alpha) N \frac{\bar{w}(N)+\bar{x}(N)}{B}\right)\bar{x}(N)\\
+&\left (1-N \bar{x}(N)-N \bar{w}(N)-\alpha \frac{N \bar{w}(N)}{\beta W}-(1-\alpha) N\frac{\bar{w}(N)+\bar{x}(N)}{B}\right)\bar{w}(N)
\\
= & \left(1 - \rho(N)-\alpha N \frac{\bar{x}(N)}{B+(1-\beta)W} -(1-\alpha)  \frac{\rho(N)}{B}\right)\bar{x}(N)\\
+&\left (1-\rho(N)-\alpha \frac{N \bar{w}(N)}{\beta W}-(1-\alpha) \frac{\rho(N)}{B}\right)\bar{w}(N)
\end{align*}
Hence, total revenue is
$$(1-\rho(N))\rho(N) - \frac{(1-\alpha)}{B}\rho^2(N) - \frac{\alpha N^2 \bar{x}(N)^2}{B + (1-\beta)W} - \frac{\alpha N^2 \bar{w}(N)^2}{\beta W}.$$
As consumer surplus is $\frac{\rho(N)^2}{2}$ it follows that total surplus is
$$\rho(N) - \frac{\rho(N)^2}{2} - \frac{(1-\alpha)}{B}\rho^2(N) - \frac{\alpha N^2 \bar{x}(N)^2}{B + (1-\beta)W} - \frac{\alpha N^2 \bar{w}(N)^2}{\beta W}.$$

Plots for various parameter values suggest that social welfare is a convex function of $\beta$ for each $N$. If true,  maximization over $\beta$ is achieved at one of the endpoints, i.e., either $0$ or $1$. The maximizer $\beta^*(N)$ is initially $0$, and it jumps to $1$ and remains there for large enough $N$. Most examples show this is between 2 and 3 (see Figure \ref{fig:SW_N}).
\subsection{Proof of Theorem \ref{limit1}}\label{prooflimit1}
Recall that the total traffic carried is given by
\begin{align*}
\rho:=x^*+w^*=\frac{B \big( B + W (1+\beta) \big)}{\big(B+2 (1-\alpha) \big) \big( B + W (1+\beta) \big)+2 B \alpha}.
\end{align*}
It is straightforward to verify that the total quantity carried, and hence, the consumer surplus are both maximized at $\beta=1$.

%{\color{red} Numerical calculations of the general case show that social welfare is always maximized at $\beta=0$. When $W$ is large, then the difference between $\beta=0$ and $\beta=1$ is small but positive nevertheless. I think this hold in general....} 

Now social welfare as a function of $\beta$ denoted $SW(\beta)$ is given by:
$$
SW(\beta)=\frac{(B+W+W\beta)^2(\frac{B^2}{2}+B(1-\alpha))+B^2\alpha(B+W-W\beta).}{\left [(B+W+W\beta)(B+2(1-\alpha))+2B\alpha\right ]^2}.
$$

%Notice that when $\alpha=0$ $SW(\beta)=\frac{B}{2(B+2)}.$ This makes sense!

The derivative of $SW(\beta)$ with respect to $\beta$ when set to zero has a unique solution, $\beta^*$. For $0\le \beta < \beta^*$ the derivative is negative, for $\beta^* < \beta \le 1$, it is positive.\footnote{ As $SW(\beta)$ is the ratio of two affine functions with a positive denominator, it is quasi-convex.} Thus to find the maximum value it is sufficient to compare the values at the two extremes. Now,
$$SW(0) = \frac{(B+W)^2(\frac{B^2}{2}+B(1-\alpha))+B^2\alpha(B+W)}{\left [(B+W)(B+2(1-\alpha))+2B\alpha\right ]^2}
$$
and
$$
SW(1) =
\frac{(B+2W)^2(\frac{B^2}{2}+B(1-\alpha))+B^3\alpha)}{\left [(B+2W)(B+2(1-\alpha))+2B\alpha\right ]^2}.
$$
Now $SW(0) > SW(1)$ implies
$$\frac{(B+W)^2(B+2(1-\alpha))+2B\alpha(B+W)}{\left [(B+W)(B+2(1-\alpha))+2B\alpha\right ]^2}
>
\frac{(B+2W)^2(B+2(1-\alpha))+2B^2\alpha)}{\left [(B+2W)(B+2(1-\alpha))+2B\alpha\right ]^2}.
$$
$$\Rightarrow\,\,
\frac{B + 2 (1-\alpha) + \frac{2B\alpha}{B+W}}{(B + 2(1-\alpha) +\frac{2B\alpha}{B+W})^2}
>
\frac{B + 2 (1-\alpha) + \frac{2B^2\alpha}{(B+2W)^2}}{(B + 2(1-\alpha) +\frac{2B\alpha}{B+2W})^2}$$
$$\Rightarrow \,\,
(B + 2(1-\alpha) +\frac{2B\alpha}{B+2W})^2
>
(B + 2(1-\alpha) +\frac{2B\alpha}{B+W})(B + 2 (1-\alpha) + \frac{2B^2\alpha}{(B+2W)^2})$$
If we let $C = \frac{B + 2(1-\alpha)}{2B\alpha}$ the expression above simplifies to
$$(C + \frac{1}{B + 2W})^2 > (C + \frac{1}{B+W})(C + \frac{B}{(B+2W)^2})$$
which is clearly true.

\subsection{Proof of Theorem \ref{prop:asym}}
\label{app:asym}

We will assume that there are $B>0$ units of always available spectrum and $W>0$ units of intermittent spectrum with availability $\alpha \in (0,1]$. We will also assume that there is one ``big" SP (labeled as 1) and $N$ ``small" ones (for $N\geq 1$) with the $i^{\mathrm{th}}$ small provider labeled as $(2, i)$. The allocation of resources is as specified below:
\begin{enumerate}
\item The big SP owns the license to $B_1$ units of always available spectrum and $W_1$ units of intermittent spectrum;
\item Each of the small SPs owns the license to $B_2/N$ units of always available spectrum and $W_2/N$ units of intermittent spectrum;
\item $\beta W$ units of intermittent spectrum is available to use by all the involved providers (big and small) as shared spectrum for $\beta \in [0,1]$.
\end{enumerate} 
We will insist that $B_1, B_2 \geq 0$ with $B_1+B_2=B$, and also that $W_1, W_2 \geq 0$ with $W_1+W_2=(1-\beta)W$. However, for ease of analysis we will assume that $B_1, B_2, W_1, W_2 > 0$ and $\beta \in (0,1)$.

We will denote the amounts served by provider $1$ as $x_1$ in licensed spectrum and $w_1$ in shared spectrum. The corresponding quantities for the $i^{\mathrm{th}}$ small provider are $x_{2,i}$ and $w_{2,i}$, respectively. Then we have the following expressions for the revenue of the providers:
\begin{align*}
R_1 & = \left(1-x_1-\sum_{j=1}^N x_{2,j} - \frac{\alpha  x_1}{B_1+W_1} - \frac{(1-\alpha) (x_1+w_1)}{B_1}- w_1 - \sum_{j=1}^N w_{2,j} \right) x_1 \\
& \quad +  \left(1-x_1-\sum_{j=1}^N x_{2,j} - \frac{\alpha  \left( w_1 + \sum_{j=1}^N w_{2,j}\right)}{\beta W} - \frac{(1-\alpha) (x_1+w_1)}{B_1}- w_1 - \sum_{j=1}^n w_{2,j}  \right) w_1 \\
R_{2,i} & =  \left(1-x_1-\sum_{j=1}^N x_{2,j} - \frac{\alpha N  x_{2,i}}{B_2+W_2} - \frac{(1-\alpha)n (x_{2,i}+w_{2,i})}{B_2}- w_1 - \sum_{j=1}^N w_{2,j} \right) x_{2,i} \\
& \quad +  \left(1-x_1-\sum_{j=1}^N x_{2,j} - \frac{\alpha   \left( w_1 + \sum_{j=1}^N w_{2,j}\right)}{\beta W} - \frac{(1-\alpha)N (x_{2,i}+w_{2,i})}{B_2}- w_1 - \sum_{j=1}^N w_{2,j}  \right) w_{2,i}
\end{align*}
Taking partial derivatives, and then setting $x_{2,i}\equiv x_2$ and $w_{2,i}\equiv w_2$ (the response of all the small SPs will be the same at equilibrium as can be argued from the symmetry of the potential function) we get the partial derivatives in the quantities as
\begin{align*}
\frac{\partial R_1}{\partial x_1} & = 1-2 x_1 \left(1+ \frac{\alpha}{B_1+W_1} + \frac{1-\alpha}{B_1} \right) -2 w_1 \left( 1+ \frac{1-\alpha}{B_1} \right) - N x_2 - N w_2 \\
\frac{\partial R_1}{\partial w_1} & = 1- 2 x_1 \left( 1+\frac{1-\alpha}{B_1}\right) - 2 w_1 \left(1+\frac{\alpha}{\beta W}+\frac{1-\alpha}{B_1} \right) - N x_2  - N w_2 \left(1+\frac{\alpha}{\beta W} \right) \\
\frac{\partial R_{2,i}}{\partial x_2} & = 1-x_1 - w_1 - N x_2 \left( 1 + \frac{1}{N} + \frac{2\alpha}{B_2+W_2} +\frac{2(1-\alpha)}{B_2}\right) - N w_2 \left(1+\frac{1}{N} + \frac{2(1-\alpha)}{B_2} \right)\\
\frac{\partial R_{2,i}}{\partial w_2} & = 1-x_1 - w_1 \left( 1+ \frac{\alpha}{\beta W}\right)-N x_2 \left(1+\frac{1}{N} + \frac{2(1-\alpha)}{B_2} \right) - N w_2 \left( 1 + \frac{1}{N} + \frac{\alpha}{\beta W} +\frac{2(1-\alpha)}{B_2}\right)
\end{align*}
The equilibrium quantities are the unique set of non-negative numbers $(x_1, w_2, x_2, w_2)$ such at
\begin{align*}
\frac{\partial R_1}{\partial x_1}\leq 0, & \frac{\partial R_1}{\partial w_1} \leq 0, & \qquad
\frac{\partial R_{2,i}}{\partial x_2} \leq 0,  & \frac{\partial R_{2,i}}{\partial w_2} \leq 0 \quad \forall i=1,\dotsc, N \\
\frac{\partial R_1}{\partial x_1} x_1= & \frac{\partial R_1}{\partial w_1} w_1 = 0, &\qquad
\frac{\partial R_{2,i}}{\partial x_2} x_2=  &\frac{\partial R_{2,i}}{\partial w_2} w_2 = 0 \quad \forall i=1,\dotsc, N
\end{align*}
Note that the inequalities give the set of the first-order conditions for maximizing the potential function, and the equations the set of complementary slackness conditions for the non-negativity constraints. 

Next we will take the limit of $N\rightarrow \infty$ where we will identify the equilibrium quantities as $x_1^*$, $w_1^*$, $x_2^*$ and $w_2^*$ with the understanding\footnote{From the uniqueness of the solutions that any limit point has to satisfy, it is easily verified that the limits exist: existence of limit points holds from compactness and uniqueness of solutions using a potential function proves the remainder.} that $\lim_{n\rightarrow\infty}(x_1, w_1, N x_2, N w_2) = (x_1^*, w_1^*, x_2^*, w_2^*)$. We will denote the limiting values of the derivatives by $\Delta_x R_1$, $\Delta_w R_1$, $\Delta_x R_2$ and $\Delta_w R_2$, respectively. Then we have
\begin{align*}
\Delta_x R_1 & = 1-2 x_1^* \left(1+ \frac{\alpha}{B_1+W_1} + \frac{1-\alpha}{B_1} \right) -2 w_1^* \left( 1+ \frac{1-\alpha}{B_1} \right) -  x_2^* - w_2^* \\
\Delta_w R_1 & = 1- 2 x_1^* \left( 1+\frac{1-\alpha}{B_1}\right) - 2 w_1^* \left(1+\frac{\alpha}{\beta W}+\frac{1-\alpha}{B_1} \right) - x_2^*  - w_2^* \left(1+\frac{\alpha}{\beta W} \right) \\
\Delta_x R_2 & = 1-x_1^* - w_1^* - x_2^* \left( 1 + \frac{2\alpha}{B_2+W_2} +\frac{2(1-\alpha)}{B_2}\right) - w_2^* \left(1 + \frac{2(1-\alpha)}{B_2} \right)\\
\Delta_w R_2 & = 1-x_1^* - w_1^* \left( 1+ \frac{\alpha}{\beta W}\right)-x_2^* \left(1 + \frac{2(1-\alpha)}{B_2} \right) - w_2^* \left( 1 + \frac{\alpha}{\beta W} +\frac{2(1-\alpha)}{B_2}\right)
\end{align*}
We will also have
\begin{align*}
\Delta_x R_1\leq 0, & \Delta_w R_1 \leq 0, & \qquad
\Delta_x R_2 \leq 0,  & \Delta_w R_2 \leq 0 \quad \forall i=1,\dotsc, n \\
\Delta_x R_1 x_1^*= & \Delta_w R_1 w_1^* = 0, &\qquad
\Delta_x R_2 x_2^*=  &\Delta_wR_2 w_2^* = 0 \quad \forall i=1,\dotsc, n
\end{align*}

Given the asymmetry between the SP 1 and the small ones, we will have to consider the possibility of SP 1 not using the shared spectrum. Using the asymptotic equilibrium quantities we will next provide\footnote{A full proof is omitted as the logic is exactly the same as in the proof of Theorem \ref{thm:asym}.} an inequality for the parameters which when satisfied will imply the existence of an $N^*$ such that for all $N\geq N^*$, in equilibrium SP 1 will abandon the shared spectrum. If the parameters are such that the inequality does not hold, then we will always have an interior point equilibrium for any $n$ but with the possibility that the limiting $w_1^*$ is zero. It is easily argued that $x_1^*$, $x_2^*$, $w_2^*$ have to be positive. 

The results can summarized as follows:
\begin{enumerate}
\item If the parameters are such that 
\begin{align}\label{eqn:asympcond1}
B_1 + W_1 + 2 (1-\alpha) \frac{W_1}{B_1} > 
%2 (1-\alpha) \frac{W_2}{B_2} + 4 (1-\alpha) \frac{\beta W}{B_2},
2 (1-\alpha) \frac{2\beta W + W_2}{B_2},
\end{align}
then there exists an $N^*$ such that for all $N \geq N^*$, SP 1 abandons the shared spectrum so that the asymptotic equilibrium quantities are $(x_1^*, w_1^*=0, x_2^*, w_2^*)$ where $(x_1^*, x_2^*, w_2^*)$ are obtained as the solution to
\begin{align}\label{eqn:asympBigSPvacate}
\begin{split}
1 & = 2 x_1^* \left(1+ \frac{\alpha}{B_1+W_1} + \frac{1-\alpha}{B_1} \right) + x_2^* + w_2^* \\
1 & = x_1^* + x_2^* \left( 1 + \frac{2\alpha}{B_2+W_2} +\frac{2(1-\alpha)}{B_2}\right) + w_2^* \left(1 + \frac{2(1-\alpha)}{B_2} \right)\\
1 & = x_1^* + x_2^* \left(1 + \frac{2(1-\alpha)}{B_2} \right) + w_2^* \left( 1 + \frac{\alpha}{\beta W} +\frac{2(1-\alpha)}{B_2}\right)
\end{split}
\end{align}
\item If, instead, we have
\begin{align}\label{eqn:asympcond2}
B_1 + W_1 + 2 (1-\alpha) \frac{W_1}{B_1} \leq 2 (1-\alpha) \frac{2\beta W + W_2}{B_2},
%2 (1-\alpha) \frac{W_2}{B_2} + 4 (1-\alpha) \frac{W_3}{B_2},
\end{align}
then for all $N$ we have an interior point equilibrium so that the asymptotic equilibrium quantities $(x_1^*, w_1^*, x_2^*, w_2^*)$ solve
\begin{align}
\begin{split}
1 & = 2 x_1^* \left(1+ \frac{\alpha}{B_1+W_1} + \frac{1-\alpha}{B_1} \right) + 2 w_1^* \left( 1+ \frac{1-\alpha}{B_1} \right) +  x_2^* + w_2^* \\
1 & = 2 x_1^* \left( 1+\frac{1-\alpha}{B_1}\right) + 2 w_1^* \left(1+\frac{\alpha}{\beta W}+\frac{1-\alpha}{B_1} \right) + x_2^*  + w_2^* \left(1+\frac{\alpha}{\beta W} \right) \\
1 & = x_1^* + w_1^* + x_2^* \left( 1 + \frac{2\alpha}{B_2+W_2} +\frac{2(1-\alpha)}{B_2}\right) + w_2^* \left(1 + \frac{2(1-\alpha)}{B_2} \right)\\
1 & = x_1^* + w_1^* \left( 1+ \frac{\alpha}{\beta W}\right)+x_2^* \left(1 + \frac{2(1-\alpha)}{B_2} \right) + w_2^* \left( 1 + \frac{\alpha}{\beta W} +\frac{2(1-\alpha)}{B_2}\right)
\end{split}
\end{align}
Note that equality in \eqref{eqn:asympcond2} implies that $w_1^*=0$ so that asymptotically SP 1 reduces the quantity served in shared spectrum to $0$.
\end{enumerate}
Note that inequality in \eqref{eqn:asympcond2} resembles the condition from Theorem \ref{thm:asym}, but with a few terms on the RHS omitted owing to many small providers assumption. 
%This is because in the current analysis some portion of licensed spectrum is also intermittent.
%}
It is easily verified at $\alpha=1$ that the big SP always vacates the shared spectrum. It is also easily verified that at $\alpha=1$, the price in the shared spectrum is $0$ in the limit (LHS-RHS of the third equation in \eqref{eqn:asympBigSPvacate} is the price) so that we get the same results as Bertrand competition.
%\\
%{\color{red}Variation of $N^*$ with the parameters needs to be investigated.}

Now consider the second scenario with asymmetric providers, and
let $x_{ij}$ and $w_{ij}$ denote the quantities in the proprietary and
shared bands, respectively, for provider $i$ in subset $j$.
The announced prices for provider $i$ in subset $j$ are
\begin{align}
p_{ij} & = 1 - \sum_{j=1}^2 \sum_i (x_{ij} + w_{ij} ) - (1-\alpha) \frac{x_{ij} + w_{ij}}{B_j / n_j}
- \alpha \frac{x_{ij}}{(B_j + W_j)/n_j}\\
p_{ij}^w & = 1 - \sum_{j=1}^2 \sum_i (x_{ij} + w_{ij} ) - 
\alpha \frac{\sum_{j=1}^2 \sum_i w_{ij}}{\beta W} - (1-\alpha) \frac{x_{ij} + w_{ij}}{B_j/n_j} .
\end{align}
and the corresponding revenue is $R_{ij} = p_{ij} x_{ij} + p_{ij}^w w_{ij}$.
In what follows we will drop the $i$ subscript since the equilibrium values will
be the same within each subset of providers.

Evaluating the first-order conditions for best response and letting $N\to \infty$ gives
\begin{align}
\label{asym_firstorder1}
\left( 1 + \frac{2(1-\alpha)}{B_j} + \frac{2\alpha}{B_j + W_j} \right) \bar{x}_j
+ \left( 1 + \frac{2(1-\alpha)}{B_j} \right) \bar{w}_j + \bar{x}_{\bar{j}} 
+ \bar{w}_{\bar{j}} & = 1 \\
\left( 1 + \frac{2(1-\alpha)}{B_j} \right) \bar{x}_j +
\left( 1 + \frac{2(1-\alpha)}{B_j} + \frac{\alpha}{\beta W} \right) \bar{w}_1
+ ( 1 + \frac{\alpha}{\beta W} ) \bar{w}_{\bar{j}} + \bar{x}_{\bar{j}} & = 1
\label{asym_firstorder2}
\end{align}
where $\bar{x}_j = n_j x_j$, $\bar{w}_j = n_j w_j$, 
$j,\bar{j} \in \{1,2\}$ and $j\neq \bar{j}$.
Note that $x_j$ and $w_j$ each tend to zero as $N\to\infty$, but $\bar{x}_j$ and $\bar{w}_j$
converge to nonnegative constants. %**need to check for consistency with Vijay's conditions**

The preceding conditions apply provided that $\bar{x}_j$ and $\bar{w}_j$ are nonnegative.
Otherwise, the providers in one of the subsets do not make use of the shared band, 
i.e., $\bar{w}_i =0$ for some $i$.
%assuming $B_j / n_j > B_j / n_j$.
In that scenario, we have $\partial R_i / \partial w_i < 0$ at $w_i=0$, which gives
\begin{equation}
\bar{x}_i \left( 1 + \frac{2(1-\alpha)}{B_i} \right) + \bar{x}_{\bar{i}} +
\bar{w}_{\bar{i}} \left( 1 + \frac{\alpha}{\beta W} \right) > 1
\label{noshare_cond1}
\end{equation}
where $\bar{x}_j$, $j=1,2$, and $\bar{w}_{\bar{i}}$ are determined from 
the three conditions \eqref{asym_firstorder1} with $j=1,2$ and
\eqref{asym_firstorder2} with $j=\bar{i}$. Combining \eqref{noshare_cond1}
with the latter conditions gives the condition in Proposition \ref{prop:asym}. 
Note that the condition resembles \eqref{eqn:asympcond2} from above, 
but now with a few terms on the LHS omitted owing to the many providers 
setting (the $2(1-\alpha)$ term is then cancelled on both sides).
In contrast to the first scenario with one large SP, here the condition
\eqref{noshare_cond1} can be satisfied for either a large ($i=1$) or small ($i=2$) SP.

\subsection{Proof of Theorem \ref{thm:existence}}
\label{sec:potential}
We show that under fairly general conditions the game with $N$ providers has a unique Nash equilibrium, and with some restrictions we also obtain a potential game. Assume there are $N$ firms and assume that prices can be negative; in equilibrium the prices will be non-negative. We divide the proof up into several cases.

 \subsubsection{Linear inverse demand and latency}
  With linear inverse demand and linear latency, $W_0$ units of intermitted secondary band set aside for unlicensed access, and assuming that firm $i\in \{1,2,\dotsc,N\}$ has $B_i$ units of always available spectrum, $W_i$ units of the intermittent secondary band, the utility of firm $i\in \{1,2,\dotsc,N\}$ is given by
\begin{align*}
u_i(y_i,w_i,\mathbf{y}_{-i}, \mathbf{w}_{-i}) & = \left(1-\sum_{j=1}^N y_j\right) y_i - (1-\alpha) \frac{y_i}{B_i} y_i - \alpha \frac{y_i-w_i}{B_i+W_i}(y_i-w_i)-\alpha \frac{\sum_{j=1}^N w_j}{W_0} w_i \\
& = y_i - y_i^2 - \sum_{j\in -i} y_j y_i - (1-\alpha) \frac{y_i^2}{B_i} - \alpha \frac{(y_i-w_i)^2}{B_i+W_i} - \alpha \frac{w_i^2}{W_0} -\alpha \frac{\sum_{j\in -i} w_j w_i}{W_0}
\end{align*}

Define the following function $\Phi(\mathbf{y},\mathbf{w})$ given by
\begin{align*}
\Phi(\mathbf{y},\mathbf{w})& = \sum_{i=1}^N y_i - \sum_{i=1}^N y_i^2 \left(1+\frac{1-\alpha}{B_i}\right)- \sum_{i=2}^N \sum_{j=1}^{i-1} y_i y_j  - \alpha \sum_{i=1}^N \frac{(y_i-w_i)^2}{B_i+W_i} \\
& \quad 
- \frac{\alpha}{W_0} \sum_{i=1}^N w_i^2 - \frac{\alpha}{W_0} \sum_{i=2}^N\sum_{j=1}^{i-1} w_i w_j
\end{align*}
Then it is easily verified that
\begin{align*}
\Phi(y_i^1,w_i^1,\mathbf{y}_{-i}, \mathbf{w}_{-i}) - \Phi(y_i^2,w_i^2,\mathbf{y}_{-i}, \mathbf{w}_{-i})=u_i(y_i^1,w_i^1,\mathbf{y}_{-i}, \mathbf{w}_{-i}) - u_i(y_i^2,w_i^2,\mathbf{y}_{-i}, \mathbf{w}_{-i})
\end{align*}
Therefore, we have a potential game. Furthermore, it is easily verified that $\Phi(\mathbf{y},\mathbf{w})$ is jointly concave in $\mathbf{y}$, $\mathbf{w}$ with the Hessian positive definite if $\alpha > 0$. If the unique maximum (under our convex and compact constraint set) also leads to non-negative prices, then it is the equilibrium. In fact, one can impose non-negative prices as constraints on the actions, and then the resulting unique maximum is a generalized equilibrium \cite{Rosen65}.

\subsubsection{Linear inverse demand and convex latency}
We can generalize the potential game characterization to the case where all providers have proprietary latency functions that are convex and (strictly) increasing. However, we still have to assume that the inverse demand function and the latency function in whitespace are both linear. A fact that we will use is the following: $l(x)$  convex and non-decreasing for $x\geq 0$ implies that $x l(x)$ is also convex, and $l(x)$ being monotonically increasing implies that $x l(x)$ is strictly convex. Again assume that $W_0$ units of the intermittent secondary band is set aside for unlicensed access.

The profit of firm $i\in \{1,2,\dotsc,N\}$ is now given by
\begin{align*}
& u_i(y_i,w_i,\mathbf{y}_{-i}, \mathbf{w}_{-i}) \\
& = \left(1-\sum_{j=1}^N y_j\right) y_i - (1-\alpha) l_i(y_i) y_i - \alpha l_{i,w}(y_i-w_i) (y_i-w_i)-\alpha \frac{\sum_{j=1}^N w_j}{W_0} w_i \\
& = y_i - y_i^2 - \sum_{j\in -i} y_j y_i - (1-\alpha) l_i(y_i) y_i - \alpha l_{i,w}(y_i-w_i) (y_i-w_i) 
%\\ & \quad 
- \alpha \frac{w_i^2}{W_0} -\alpha \frac{\sum_{j\in -i} w_j w_i}{W_0}
\end{align*}
Then, the potential function $\Phi(\mathbf{y},\mathbf{w})$ is given by
\begin{align*}
\Phi(\mathbf{y},\mathbf{w})& = \sum_{i=1}^N y_i - \sum_{i=1}^N y_i^2 - \sum_{i=2}^N \sum_{j=1}^{i-1} y_i y_j  - (1-\alpha) \sum_{i=1}^N l_i(y_i) y_i - \alpha \sum_{i=1}^N l_{i,w}(y_i-w_i) (y_i-w_i) \\
& \quad 
- \frac{\alpha}{W_0} \sum_{i=1}^N w_i^2 - \frac{\alpha}{W_0} \sum_{i=2}^N\sum_{j=1}^{i-1} w_i w_j
\end{align*}
If the inverse demand function is $P(y)=1-\gamma y$ for some $\gamma>0$, then, the potential function $\Phi(\mathbf{y},\mathbf{w})$ is given by
\begin{align*}
\Phi(\mathbf{y},\mathbf{w})& = \sum_{i=1}^N y_i - \gamma \sum_{i=1}^N y_i^2 - \gamma \sum_{i=2}^N \sum_{j=1}^{i-1} y_i y_j  - (1-\alpha) \sum_{i=1}^N l_i(y_i) y_i - \alpha \sum_{i=1}^N l_{i,w}(y_i-w_i) (y_i-w_i) \\
& \quad 
- \frac{\alpha}{W_0} \sum_{i=1}^N w_i^2 - \frac{\alpha}{W_0} \sum_{i=2}^N\sum_{j=1}^{i-1} w_i w_j
\end{align*}
 \subsubsection{Concave inverse demand and convex latency}
Finally, we consider the existence of pure equilibria in the general case where, the inverse demand  is a general concave decreasing function $P(\cdot)$ and the latency function in whitespace is a general convex increasing function $l_w(\cdot)$. Assuming the latency cost to be a function of the normalized load\footnote{These correspond to the latency cost for proprietary spectrum of provider $i$ being $\ell_i(x)=f_i(x/B_i)$ for some $B_i>0$ and $\ell_{i,w}(x)=f_i(x/(B_i+W_i))$ for some $W_i \geq 0$ with $f_i(\cdot)$ convex and increasing, and $l_w(w)=f_w(w/W)$ for some $W>0$ and $f_w(\cdot)$ convex and increasing.} incorporating the capacity provisioned is a special case of our general setting. In this case the utility of firm $i\in \{1,2,\dotsc,N\}$ is now given by
\begin{align*}
u_i(y_i,w_i,\mathbf{y}_{-i}, \mathbf{w}_{-i}) & = y_i P\left(y_i + \sum_{j\in -i} y_j\right) - (1-\alpha) l_i(y_i) y_i \\
& \quad - \alpha l_{i,w}(y_i-w_i) (y_i-w_i) - \alpha w_i l_w\left(w_i + \sum_{j\in -i} w_j\right),
\end{align*}
where $-i := \{1,2,\dotsc,N\}\setminus \{i\}$.
It is easily verified that given the strategy of the opponents, namely $(\mathbf{y}_{-i}, \mathbf{w}_{-i})$, the utility of firm $i$ is jointly concave in $(y_i, w_i)$, and where $(\mathbf{y},\mathbf{w})$ are to be chosen from a compact and convex set\footnote{The constraints are $\sum_i y_i \leq 1$, $y_i \geq w_i \geq 0$ for all $i\in \{1, 2, \dotsc, N\}$ and the prices being non-negative.}.%which are to be chosen in a convex and compact set with the additional linear constraint that $y_i\geq w_i$. 
Therefore, we have a concave game and existence of pure equilibria follows from the results of \cite{Rosen65}.

Following up regarding the uniqueness of equilibria, using \cite{Rosen65} (taking $r_i\equiv 1$ for all $i=1, 2, \dotsc, N$) and working with variables $x_i=y_i-w_i$ and $w_i$, we need to determine the Jacobian $G$ of the gradient vector $g$ and show that $H=G+G^T$ is negative definite, where for $i=1, 2, \dotsc, N$, 
\begin{align*}
g_{x_i} & =\frac{\partial u_i(x_i,w_i,\mathbf{x}_{-i},\mathbf{w}_{-i})}{\partial x_i} \\
& =P(\sum_{j=1}^N x_j + w_j) + (x_i+w_i) P^\prime(\sum_{j=1}^N x_j + w_j) - (1-\alpha) ( l_i(x_i+w_i)+(x_i+w_i) l_i^\prime(x_i+w_i)) \\
& \qquad - \alpha ( l_{i,w}(x_i) + x_i l_{i,w}^\prime(x_i))\\
g_{w_i} & = \frac{\partial u_i(x_i,w_i,\mathbf{x}_{-i},\mathbf{w}_{-i})}{\partial w_i} \\
& = P(\sum_{j=1}^N x_j + w_j) + (x_i+w_i) P^\prime(\sum_{j=1}^N x_j + w_j) - (1-\alpha) ( l_i(x_i+w_i)+(x_i+w_i) l_i^\prime(x_i+w_i)) \\
& \qquad - \alpha ( l_w(\sum_{j=1}^N w_j) + w_i l_w^\prime(\sum_{j=1}^N w_j))
\end{align*}
where we use the original strategy space $(x_i,w_i)$ for each of the providers and label each component by the corresponding variable. Note that for $i, j \in \{1, 2, \dotsc, N\}$,
\begin{align*}
G_{x_i,x_j} & = \frac{\partial^2 u_i(x_i,w_i,\mathbf{x}_{-i},\mathbf{w}_{-i})}{\partial x_i \partial x_j},  &
G_{x_i,w_j} & = \frac{\partial^2 u_i(x_i,w_i,\mathbf{x}_{-i},\mathbf{w}_{-i})}{\partial x_i \partial w_j},\\
G_{w_i,x_j} & = \frac{\partial^2 u_i(x_i,w_i,\mathbf{x}_{-i},\mathbf{w}_{-i})}{\partial w_i \partial x_j},  & 
G_{w_i,w_j}  & = \frac{\partial^2 u_i(x_i,w_i,\mathbf{x}_{-i},\mathbf{w}_{-i})}{\partial w_i \partial w_j} 
\end{align*}
and
\begin{align*}
H_{x_i,x_j}=H_{x_j,x_i} & = \frac{\partial^2 u_i(x_i,w_i,\mathbf{x}_{-i},\mathbf{w}_{-i})}{\partial x_i \partial x_j} + \frac{\partial^2 u_j(x_i,w_i,\mathbf{x}_{-i},\mathbf{w}_{-i})}{\partial x_i \partial x_j}, \\ 
H_{x_i,w_j}=H_{w_j,x_i}&=\frac{\partial^2 u_i(x_i,w_i,\mathbf{x}_{-i},\mathbf{w}_{-i})}{\partial x_i \partial w_j}+\frac{\partial^2 u_j(x_i,w_i,\mathbf{x}_{-i},\mathbf{w}_{-i})}{\partial x_i \partial w_j} \\
H_{w_i, x_j}=H_{x_j,w_i} & = \frac{\partial^2 u_i(x_i,w_i,\mathbf{x}_{-i},\mathbf{w}_{-i})}{\partial w_i \partial x_j} + \frac{\partial^2 u_j(x_i,w_i,\mathbf{x}_{-i},\mathbf{w}_{-i})}{\partial w_i \partial x_j} \\
H_{w_i,w_j}=H_{w_j,w_i} & = \frac{\partial^2 u_i(x_i,w_i,\mathbf{x}_{-i},\mathbf{w}_{-i})}{\partial w_i \partial w_j} + \frac{\partial^2 u_j(x_i,w_i,\mathbf{x}_{-i},\mathbf{w}_{-i})}{\partial w_i \partial w_j}
\end{align*}
We have the following for $i=1, 2, \dotsc, N$,
\begin{align*}
\frac{\partial^2 u_i(x_i,w_i,\mathbf{x}_{-i},\mathbf{w}_{-i})}{\partial x_i \partial x_i} & = 2 P^\prime(\sum_{k=1}^N x_k + w_k) + (x_i+w_i) P^{\prime\prime}(\sum_{k=1}^N x_k + w_k) \\
& \qquad - (1-\alpha) ( 2 l_i^\prime(x_i+w_i)+(x_i+w_i) l_i^{\prime\prime}(x_i+w_i) ) \\
& \qquad - \alpha ( 2 l_{i,w}^\prime(x_i) + x_i l_{i,w}^{\prime\prime}(x_i) ) \\
\frac{\partial^2 u_i(x_i,w_i,\mathbf{x}_{-i},\mathbf{w}_{-i})}{\partial w_i \partial w_i} & = 2 P^\prime(\sum_{k=1}^N x_k + w_k) + (x_i+w_i) P^{\prime\prime}(\sum_{k=1}^N x_k + w_k) \\
& \qquad - (1-\alpha) ( 2 l_i^\prime(x_i+w_i)+(x_i+w_i) l_i^{\prime\prime}(x_i+w_i) ) \\
& \qquad - \alpha ( 2 l_w^\prime(\sum_{k=1}^N w_k) + w_i l_w^{\prime\prime}(\sum_{k=1}^N w_k) ) \\
\frac{\partial^2 u_i(x_i,w_i,\mathbf{x}_{-i},\mathbf{w}_{-i})}{\partial x_i \partial w_i} & = 2 P^\prime(\sum_{k=1}^N x_k + w_k) + (x_i+w_i) P^{\prime\prime}(\sum_{k=1}^N x_k + w_k) \\
& \qquad - (1-\alpha) ( 2 l_i^\prime(x_i+w_i)+(x_i+w_i) l_i^{\prime\prime}(x_i+w_i) ) 
\end{align*}
For $i, j \in \{1, 2, \dotsc, N\}$ with $i\neq j$ we have
\begin{align*}
\frac{\partial^2 u_i(x_i,w_i,\mathbf{x}_{-i},\mathbf{w}_{-i})}{\partial x_i \partial x_j} & = P^\prime(\sum_{k=1}^N x_k + w_k) + (x_i+w_i) P^{\prime\prime}(\sum_{k=1}^N x_k + w_k) \\
\frac{\partial^2 u_i(x_i,w_i,\mathbf{x}_{-i},\mathbf{w}_{-i})}{\partial x_i \partial w_j} & = P^\prime(\sum_{k=1}^N x_k + w_k) + (x_i+w_i) P^{\prime\prime}(\sum_{k=1}^N x_k + w_k) \\
\frac{\partial^2 u_i(x_i,w_i,\mathbf{x}_{-i},\mathbf{w}_{-i})}{\partial w_i \partial x_j} & = P^\prime(\sum_{k=1}^N x_k + w_k) + (x_i+w_i) P^{\prime\prime}(\sum_{k=1}^N x_k + w_k) \\
\frac{\partial^2 u_i(x_i,w_i,\mathbf{x}_{-i},\mathbf{w}_{-i})}{\partial w_i \partial w_j} & = P^\prime(\sum_{k=1}^N x_k + w_k) + (x_i+w_i) P^{\prime\prime}(\sum_{k=1}^N x_k + w_k) \\
& \qquad - \alpha ( l_w^\prime(\sum_{k=1}^N w_k) + w_i l_w^{\prime\prime}(\sum_{k=1}^N w_k) )
\end{align*}
Therefore, $z^T H z$ is given by
\begin{align*}
\mathbf{z}^T H \mathbf{z} & = 2 P^\prime\left((\mathbf{x}+\mathbf{w})^T\mathbf{1}\right) \left[ (\mathbf{z}^T \mathbf{1})^2 + (\mathbf{z_x}+\mathbf{z_w})^T (\mathbf{z_x}+\mathbf{z_w}) \right] \\
& \quad + 2 P^{\prime\prime} \left((\mathbf{x}+\mathbf{w})^T\mathbf{1}\right) (\mathbf{z_x}+\mathbf{z_w})^T \mathbf{1} (\mathbf{x}+\mathbf{w})^T (\mathbf{z_x}+\mathbf{z_w}) \\
& \quad - \alpha \sum_{i=1}^N \left( 2 l_{i,w}^\prime(x_i) + x_i l_{i,w}^{\prime\prime}(x_i) \right) z_{x_i}^2 \\
& \quad  - (1-\alpha) \sum_{i=1}^N \left(2 l_i^\prime(x_i+w_i)+(x_i+w_i) l_i^{\prime\prime}(x_i+w_i) \right) (z_{x_i} + z_{w_i})^2 \\
& \quad - 2 \alpha l_w^\prime\left(\mathbf{w}^T\mathbf{1}\right) \left[ (\mathbf{z_w}^T \mathbf{1})^2 +\mathbf{z_w}^T \mathbf{z_w}\right]  - 2 \alpha l_w^{\prime\prime}\left( \mathbf{w}^T\mathbf{1} \right)\mathbf{z_w}^T \mathbf{1} \mathbf{w}^T \mathbf{z_w}
\end{align*}
If either $P^\prime(0)<0$ and $l_w^\prime(0)>0$ or $l_i^{\prime}(0)>0$ and $l_{i,w}^\prime(0)>0$ for all $i=1, 2, \dotsc, N$, then it follows that $H$ is negative definite, where we've also used the fact that $\mathbf{1} (\mathbf{x}+\mathbf{w})^T$ and $\mathbf{1}\mathbf{w}^T$ are positive semidefinite since $\mathbf{x}$ and $\mathbf{w}$ are non-negative vectors. Under these conditions we have a unique equilibrium. Note that $P^\prime(0)<0$ and $l_w^\prime(0)>0$ is a sufficient condition for $xP(x)$ being strictly concave and $x l_w(x)$ being strictly convex, and similarly, $l_i^{\prime}(0)>0$ and $l_{i,w}^\prime(0)>0$ for all $i=1, 2, \dotsc, N$ is a sufficient condition for $x l_i(x)$ and $x l_{i,w}(x)$ being strictly convex for all $i=1,2,\dotsc, N$.

The case of no shared spectrum is equivalent to $\alpha=0$ above with the understanding that provider $i$ only chooses $y_i=x_i+w_i$. In that case (with dimension of $H$ being $N\times N$) we have
\begin{align*}
\mathbf{z}^T H \mathbf{z} & = 2 P^\prime\left(\mathbf{y}^T\mathbf{1}\right) \left[ (\mathbf{z}^T \mathbf{1})^2 + \mathbf{z}^T\mathbf{z} \right]  + 2 P^{\prime\prime} \left(\mathbf{y}^T\mathbf{1}\right) \mathbf{z}^T \mathbf{1} \mathbf{y}^T \mathbf{z} \\
& \quad - \sum_{i=1}^N \left( 2 l_i^\prime(y_i) + y_i l_i^{\prime\prime}(y_i) \right) z_{y_i}^2
\end{align*}
If either $P^\prime(0)<0$ or $l_i^{\prime}(0)>0$ for all $i=1, 2, \dotsc, N$, then it follows that $H$ is negative definite, and we have a unique equilibrium.

\textit{Generalizations:} The existence and uniqueness of the Nash equilibria also extends to more general availability scenarios, with a similar proofs. For example, we can allow the proprietary spectrum of different providers to also have a general distribution of availability with possibly multiple bands with the only restriction being that every service provider always has a minimum non-zero amount of spectrum available for proprietary use. Similarly, we can also allow multiple shared bands and also the proprietary spectrum with a general distribution, with the restriction that whenever the total amount of shared bandwidth is zero, there is non-zero proprietary bandwidth available at every provider and every service provider always has a minimum non-zero amount of spectrum available for proprietary use.

\subsubsection{Structure}\label{structure}
%Asymmetric Unlicensed Regime for  $N=2$}
\label{app:structure}

%{\color{red} Vijay: needs editing.}

Theorem \ref{thm:existence} gives us a unique equilibrium. It is easy to see that prices being $0$ at equilibrium can only occur if the quantity is also zero: if not, then reducing the quantity by $\epsilon$ leads to non-zero price and an increase in profit. The unique equilibrium also maximizes a concave potential function when the demand function is linear and all the latencies are linear too. We use this and the KKT theorem to characterize the structure of the equilibrium. We establish the following results: 
\begin{enumerate}
\item Considering the case of $N=2$ in Result 1 we provide necessary and sufficient conditions for a provider to not use the shared spectrum in equilibrium. 
\item Again specializing to $N=2$, in Result 2 we show that the proprietary spectrum bands are always used in equilibrium. We also show that the logic extends to $N>2$ also.
\item In Result 3 we provide the counterpoint to Result 1 to determine necessary and sufficient conditions for both providers to use all available spectrum bands.
\item In Result 4 we generalize Result 1 to the case of $N>2$ and provide necessary and sufficient conditions for all but one provider to not use the shared spectrum in equilibrium.
\item In Result 5 we further generalize Results 1 through 4 to the case when some of the licensed bands are also intermittent.
\end{enumerate}

\textbf{Result 1}: The equilibrium is such that $w_i^*=0$, $y_i^*=x_i^*>0$, and $y_{-i}^* > w_{-i}^* > 0$ if and only if 
$B_i \geq 2 W + 4(1-\alpha) \frac{W}{B_{-i}}+2 B_{-i} +2$, so that provider $i$ does not use the whitespace spectrum while provider $-i$ gets proprietary access.\\
\begin{proof}
Note that $N=2$. Let $i=1$ wlog so that $-i=2$. Then we have
\begin{align*}
\frac{\partial\Phi}{\partial y_1}&=1-2y_1\left(1+\frac{1-\alpha}{B_1} \right) -y_2 - \frac{2\alpha}{B_1}(y_1-w_1) \\
\frac{\partial\Phi}{\partial y_2}&=1-2y_2\left(1+\frac{1-\alpha}{B_2} \right) -y_1 - \frac{2\alpha}{B_2}(y_2-w_2) \\
\frac{\partial\Phi}{\partial w_1}&=-\frac{2\alpha}{B_1} (w_1-y_1) - \frac{2\alpha}{W} w_1 - \frac{\alpha}{W} w_2 \\
\frac{\partial\Phi}{\partial w_2}&=-\frac{2\alpha}{B_2} (w_2-y_2) - \frac{2\alpha}{W} w_2 - \frac{\alpha}{W} w_1
\end{align*}
If the optimum (and also equilibrium) is such that $w_1^*=0$, $y_1^*=x_1^*>0$, and $y_2^* > w_2^* > 0$, then $\tfrac{\partial\Phi}{\partial y_1}=0$, $\tfrac{\partial\Phi}{\partial y_2}=0$, $\tfrac{\partial\Phi}{\partial w_2}=0$ and $\tfrac{\partial\Phi}{\partial w_1}\leq 0$ at the optimum. Substituting the variables and using the above constraints, we get
\begin{align*}
0& =1-2y_1^*\left(1+\frac{1-\alpha}{B_1} \right) -y_2^* - \frac{2\alpha}{B_1}y_1^* \\
0 & =1-2y_2^*\left(1+\frac{1-\alpha}{B_2} \right) -y_1^* - \frac{2\alpha}{B_2}(y_2^*-w_2^*) \\
0 &=-\frac{2\alpha}{B_2} (w_2^*-y_2^*) - \frac{2\alpha}{W} w_2^*  \\
0 & \geq \frac{2\alpha}{B_1} y_1^*  - \frac{\alpha}{W} w_2^*
\end{align*}
The third equation implies that 
\begin{align*}
y_2^*=\left(1+\frac{B_2}{W}\right) w_2^* > w_2^* \quad (\text{If } w_2^*> 0)
\end{align*}
Using this we get two equations in two unknowns, $y_1^*$ and $w_2^*$. Solving these yields
\begin{align*}
w_2^* & = \frac{1+\frac{2}{B_1}}{(4(1+\frac{1}{B_1})(1+\frac{1}{B_2})-1)(1+\frac{B_2}{W})-\frac{4\alpha}{B_2}(1+\frac{1}{B_1})} \\
y_1^* & = \frac{1+\frac{2(1-\alpha)}{B_2}+\frac{B_2}{W}+\frac{2}{W}}{(4(1+\frac{1}{B_1})(1+\frac{1}{B_2})-1)(1+\frac{B_2}{W})-\frac{4\alpha}{B_2}(1+\frac{1}{B_1})} 
\end{align*}
Note that both are positive. With some algebra it is also verified that all the prices are non-negative. Thus, the last condition that must hold is $\tfrac{\partial\Phi}{\partial w_1}\leq 0$, which implies that $\tfrac{2}{B_1} y_1^*  \leq \tfrac{1}{W} w_2^*$, i.e., 
\begin{align*}
B_1 \geq 2 W + 4(1-\alpha) \frac{W}{B_2}+2 B_2 +2
\end{align*}
Additionally, it can also be verified for all $B_1$ satisfying the above inequality
\begin{align*}
B_2 < 2 W + 4(1-\alpha) \frac{W}{B_1}+2 B_1 +2
\end{align*}
This proves the result. \hfill $\Box$
\end{proof}\\
\textit{Remarks:}
\begin{enumerate}
\item The proof above holds by checking for conditions when the chosen equilibrium maximizes the potential function; the condition is equivalent to the partial derivative of the potential function in $w_i$ being non-positive. For fixed $B_{-i}$, $W$ and $\alpha$ for all $B_i$ sufficiently large, the condition above will hold. Keeping $B_i$, $W$ and $\alpha$ fixed such that $B_i \geq 2 W+2+4\sqrt{2(1-\alpha)W}$ (RHS is minimum value of lower bound as $B_{-i}$ is varied), there are two values $B_{-i}^{lb}$ and $B_{-i}^{ub}$ (corresponding to solution to quadratic with equality in constraint) such that if $B_{-i} \in [B_{-i}^{lb}, B_{-i}^{ub}]$, then provider $i$ vacates the share spectrum, and otherwise she uses it; if $B_i < W+2+4\sqrt{2(1-\alpha)W}$, then provider $i$ always uses the shared spectrum. 
\item From constrained optimization theory we know that at the equilibrium we will have $\tfrac{\partial\Phi}{\partial w_1} \leq 0$ and $\tfrac{\partial\Phi}{\partial w_2} \leq 0$. Rewriting these in terms of the equilibrium variables (and assuming $\alpha>0$) we get
\begin{align*}
\frac{x_i^*}{B_i} \leq \frac{w_i^*+\frac{w_{-i}^*}{2}}{W} \leq \frac{w_1^*+w_{2}^*}{W}, 
\end{align*}
which is equivalent to stating that the congestion level in the proprietary bands is always less than the congestion level in the shared band; note that if the provider $i$ uses the shared band, then the first inequality is tight. Additionally, if in equilibrium both providers carry non-zero traffic in the shared band, then the congestion level in the shared band is strictly greater, by exactly $\tfrac{w_{-i}^*}{2W}$ for provider $i\in\{1,2\}$. Furthermore, if in equilibrium provider $i\in \{1,2\}$ does not use the shared band, then the proprietary band for provider $-i$ and the shared band have the same congestion level that is at least two times the level of the congestion in provider $i$'s proprietary band. Interestingly, this is the only reason for provider $i$ to not use the shared band, as opposed to others such as the (corresponding) price becoming negative if the traffic carried is positive, etc.
\end{enumerate}

%Unclear how this result generalizes to more than two providers: will two out of three vacate whitespace or will there always be competition in whitespace when $N>2$?
%Numerical investigations when $N>2$ show the following. Assume that the capacities of the providers are in decreasing order. Then there exists $i^*\in \{1, 2, \dotsc, N\}$ such that only providers $\{i^*, \dotsc, N\}$ use whitespace spectrum and the remaining providers (if any) do not use it. If $B_i \geq 2 W + 4 (1-\alpha) \tfrac{W}{B_N} + 2 B_N +2$ for all $i\in \{1, 2, \dotsc, N-1\}$, then $i^*=N$ so that provider $N$ is the sole user of whitespace spectrum! If $B_{N-1}$ and $B_N$ are sufficiently close, then it is possible that $i^*=N-1$ even when $B_i < \min(2 W + 4 (1-\alpha) \tfrac{W}{B_N} + 2 B_N +2, 2 W + 4 (1-\alpha) \tfrac{W}{B_{N-1}} + 2 B_{N-1} +2)$ for all $i\in \{1, 2, \dotsc, N-2\}$.

\textbf{Result 2}: At the equilibrium $x_i^*>0$ for all $i=1,2$, irrespective of the parameters. \\
\begin{proof}
As before let $i=1$ wlog so that $-i=2$. If $x_1^*=0$, then $y_1^*=w_1^*>0$ (easy to see that at least one of $x_1^*$ or $w_1^*$ should be positive). This then implies that 
\begin{align*}
\frac{\partial\Phi}{\partial y_1} \leq 0, \text{ and } \frac{\partial\Phi}{\partial w_1} \geq 0.
\end{align*}
The second inequality can be rewritten as 
\begin{align*}
- \frac{2\alpha}{W} w_1^* - \frac{\alpha}{W} w_2^* \geq 0,
\end{align*}
which then implies that $w_1^*=w_2^*=0$. This is a contradiction. \hfill $\Box$ %\qedsymbol
\end{proof}\\
Using the same logic, this result holds for $N>2$ too. For the $N=2$ case, we next show that Result 2 implies Result 3.

\textbf{Result 3}: For $N=2$, if the conditions of Result 1 don't hold, then the equilibrium is always an interior point equilibrium. \\
\begin{proof}
From Result 2 we know that $x_i^*>0$ for all $i=1,2$. Since the conditions of Result 1 don't apply, we can either have $w_1^*=w_2^*=0$ or $w_1^*, w_2^*>0$. The former cannot hold as this would imply 
\begin{align*}
\frac{\partial\Phi}{\partial w_i} \leq 0 
\Leftrightarrow \frac{2\alpha}{B_i} x_i^* \leq 0 
\Leftrightarrow x_i^*=0,
\end{align*}
which is a contradiction. \hfill $\Box$
\end{proof}

%Result 1 generalizes easily to the following.

\textbf{Result 4}: For general $N$, the equilibrium is $y_i^*=x_i^*>0$, $w_i^*=0$ for all $i=1,2,\dotsc,N-1$ and  $y_N^*> w_N^*>0$ if and only if $B_i \geq 2 W + 2B_N + 2 + 4 (1-\alpha) \tfrac{W}{B_N}$ for all $i=1,2,\dotsc,N-1$. Note that all providers except for provider $N$ vacate the whitespace spectrum\\
\begin{proof}
Using the potential function and the KKT conditions for convex optimization, the equilibrium is $y_i^*=x_i^*>0$, $w_i^*=0$ for all $i=1,2,\dotsc,N-1$ and  $y_N^*> w_N^*>0$ if and only if the following hold at the equilibrium
\begin{align}
\frac{\partial \Phi}{\partial y_i} & = 0, \quad i=1, 2, \dotsc, N \label{eq:eq1}\\
\frac{\partial \Phi}{\partial w_i} & \leq 0, \quad i=1, 2, \dotsc, N-1 \label{eq:ineq1}\\
\frac{\partial \Phi}{\partial w_N} & = 0 \label{eq:eq2}
\end{align}
Equation \eqref{eq:eq2} yields
\begin{align}
\frac{\partial \Phi}{\partial w_N} = -\frac{2\alpha}{B_N} (w_N-y_N) - \frac{2\alpha}{W} w_N = 0 \Leftrightarrow y_N^* = \left(1+\frac{B_N}{W} \right) w_N^* \label{eq:ynwn}
\end{align}
For $i=1,2, \dotsc, N-1$, inequalities \eqref{eq:ineq1} yield 
\begin{align}
\frac{\partial \Phi}{\partial w_i} = \frac{2y_i}{B_i} - \frac{w_N}{W} \leq 0 \Leftrightarrow \frac{2y_i^*}{B_i} \leq \frac{w_N^*}{W} \label{eq:yilwn}
\end{align}
For $i=N$, equation \eqref{eq:eq1} in combination with \eqref{eq:ynwn} yields
\begin{align}
\begin{split}
& \frac{\partial \Phi}{\partial y_i} = 1- 2 y_N \left(1+\frac{1}{B_N} \right) - \sum_{j=1}^{N-1} y_j + \frac{2\alpha w_N}{B_N} = 0 \\
\Leftrightarrow & \sum_{j=1}^{N-1} y_j^* = 1- 2 w_N^* \left[ \left(1+\frac{B_N}{W} \right)\left(1+\frac{1}{B_N}\right) - \frac{\alpha}{B_N}\right] 
\end{split}
\label{eq:ysumwn}
\end{align}
For $i=1,2,\dotsc,N-1$, equations \eqref{eq:eq1} in combination with \eqref{eq:ynwn} and \eqref{eq:ysumwn} yield
\begin{align}
\begin{split}
& \frac{\partial \Phi}{\partial y_i} = 1 - \sum_{j=1, j\neq i}^{N} y_j - 2 y_i \left(1+\frac{1}{B_i}\right) = 0 \\
\Leftrightarrow & y_i^* \left(1+\frac{2}{B_i} \right) = w_N^* \left[ \left(1+\frac{B_N}{W} \right) \left(1+\frac{2}{B_N}\right) - \frac{2\alpha}{B_N}\right]
\end{split}
\label{eq:yiwn}
\end{align}
Substituting the result of \eqref{eq:yiwn} into \eqref{eq:ysumwn} yields
\begin{align}
w_N^* = \frac{1}{2 \left(1+\frac{B_N}{W}\right) \left(1+\frac{1}{B_N}\right) - \frac{2\alpha}{B_N} +\left[ \left(1+\frac{B_N}{W}\right) \left(1+\frac{2}{B_N}\right)-\frac{2\alpha}{B_N}\right]\sum_{j=1}^{N-1} \frac{1}{1+\frac{2}{B_j}}} > 0
\end{align}
With some algebra it can be shown that all the prices are all positive as well. Therefore, the only conditions that need to be satisfied are given by \eqref{eq:yilwn}, which simplify to
\begin{align*}
B_i \geq 2 W + 2 B_N + 2 +4 (1-\alpha) \frac{W}{B_N}, \quad \forall i=1,2,\dotsc, N-1.
\end{align*}
This proves the result. \hfill $\Box$
\end{proof}
%\\
%Numerical investigations confirm all these results.

\textbf{Result 5:} Let the $W$ units of secondary band be split into $(W_0, W_1, \dotsc, W_N)$ such that firm $i$ gets $W_i$ units and $W_0$ units are assigned to shared access, then the following are true:
\begin{enumerate}
\item For all $N\geq 2$ the equilibrium is such that $x_i^*>0$ for all $i=1, 2, \dotsc, N$.
\item The equilibrium is $y_i^*=x_i^*>0, w_i^*=0$ for all $i=1,2, \dotsc, N-1$ and $y_N^* > w_N^* > 0$ if and only if for all $i=1,2,\dotsc, N-1$
\begin{align}\label{eq:GenVac}
B_i + W_i + 2 (1-\alpha) \frac{W_i}{B_i} \geq 2 (W_0+W_N) + 2 B_N + 2 + 4 (1-\alpha) \frac{W_0+W_N}{B_N} .
\end{align}
\item For $N=2$, if the condition in \eqref{eq:GenVac} is not satisfied for either firm $1$ or $2$, then we necessarily have an interior point equilibrium, i.e., $y_1^* > w_1^* > 0$ and $y_2^* > w_2^* > 0$.
\end{enumerate}
%{\color{red} Need to check proof as I suspect that the term $4 (1-\alpha) \tfrac{W_0+W_N}{B_N}$ is actually $2 (1-\alpha) \tfrac{2 W_0+W_N}{B_N}$.}
\begin{proof}
The proof follows by repeating all the steps of the proofs of Results 1 to 4, and is omitted. \hfill $\Box$ 
\end{proof}\\
\textit{Remark:} Similar to Remark 2 at the bottom of Result 1, at equilibrium we need $\tfrac{\partial\phi}{\partial w_i}\leq 0$ and this implies that for all $i=1, \dotsc, N$ 
\begin{align*}
\frac{x_i^*}{B_i+W_i} \leq \frac{w_i^* + \frac{\sum_{j\in -i} w_j^*}{2}}{W_0} \leq \frac{\sum_{j=1}^N w_j^*}{W_0}.
\end{align*}
This again implies that conditioned on the intermittent spectrum being available, the congestion in the proprietary band is less than the congestion in the shared band. Since $w_i^*>0$ implies that the first inequality is tight, the congestion level in the proprietary band of provider $i$ is exactly lower by $\tfrac{\sum_{j\in -i} w_j^*}{2W_0}$ and strictly greater than \textit{half} the congestion level in the shared band. If instead, $w_i^*=0$, then the congestion level in the propriety band of provider $i$ is less than \textit{half} the congestion level in the shared band. Both these statements hold when conditioned on the intermittent spectrum being available.

\subsection{Proof of Proposition \ref{marg:prop1} }
\label{app:marginal}

Given inverse demand $P(\cdot)$ and latency $\ell (\cdot)$,
the revenue for provider $k$ is
\begin{equation}
R_k = x_k [ P(x_1 + x_2 ) - \ell (x_k / B_k )] .
\end{equation}
where $B_k$ is provider $k$'s bandwidth.
Setting $\partial R_k / \partial x_k =0$ gives
\begin{equation}
P(x_1 + x_2 ) - \ell (x_k/B_k) + x_k [ P' (x_1 + x_2) - \ell ' (x_k/B_k) B_k^{-1} ] =0
\label{eq:br}
\end{equation}
Taking $\partial / \partial B_k$ gives
\begin{equation}
\frac{\partial x_k}{\partial B_k} G(x_k,x_{-k}) =
- \frac{2x_k}{B_k^2} \ell' (x_k/B_k) - \frac{x_k^2}{B_k^3} \ell'' (x_k/B_k)
\end{equation}
where
\begin{equation}
G(x_k,x_{-k}) = 
2 P' (x_1+x_2) + x_k P'' (x_1+x_2) - \frac{2}{B_k} \ell' (x_k/B_k) 
- \frac1{B_k^2} \ell'' (x_k/B_k) ,
\end{equation}
and is negative. The right-hand side is also negative, hence ${\partial x_k}/{\partial B_k} > 0$.

The best response condition \eqref{eq:br} gives $x_{k}$ implicitly as a function of $x_{-k}$.
Swapping $k$ and $-k$, we can compute
\begin{equation}
\frac{\partial x_{-k}}{\partial x_{k}} G(x_{-k}, x_k ) = - P'(x_1+x_2) - x_{-k} P'' (x_1+x_2)
\end{equation}
Since $G<0$ and the right-hand side is positive, ${\partial x_{-k}}/{\partial x_{k}} < 0$.
Hence a marginal increase in $B_k$ leads to a marginal increase in $x_k$ and a
marginal decrease in $x_{-k}$.

Now let $x_{k,-k} = \partial x_k / \partial x_{-k}$ and consider 
the sequence of best responses for $n=1,2,\cdots$, 
which result from adding $\delta B_k$ starting from
the equilibrium $(x_k^{(0)},x_{-k}^{(0)})$:
\begin{enumerate}
\item Update $x_k^{(1)} = x_k^{(0)} + \delta x_k^{(1)}$, where
$\delta x_k^{(1)} = ( \partial x_k / \partial B_k ) \cdot \delta B_k > 0$.
\item Update $x_{-k}^{(n)} = x_{-k}^{(n-1)} + \delta x_{-k}^{(n)}$
where $\delta x_{-k}^{(n)} = x_{-k,k} \cdot \delta x_k^{(n)} < 0$.
\item Update $x_k^{(n+1)} = x_k^{(n)} + \delta x_{k}^{(n+1)}$
where $\delta x_{-k}^{(n+1)} = x_{k,-k} \cdot \delta x_{-k}^{(n)} > 0$
\item Iterate steps 2 and 3.
\end{enumerate}

The total change in $x_k$ is then the geometric series
$\delta x_k (1 + \rho + \rho^2 + \cdots)$ where $\rho = x_{k,-k} x_{-k,k}$.
Since $0< \rho <1$, this sequence of best responses converges
to a new equilibrium with quantities
$(x_k + \delta x_{k,eq}, x_{-k} + \delta x_{-k,eq})$ where
\begin{align}
\delta x_{k,eq} & = \frac{1}{1 - \rho} \cdot \delta x_{k}^{(1)} > 0\\
\delta x_{-k,eq} & = x_{-k,k} \cdot \delta x_{k,eq} < 0.
\end{align}
Furthermore, the change in total quantity
\begin{equation}
\delta x_{k,eq} + \delta x_{-k,eq} = (1 + x_{-k,k}) \frac1{1-\rho} \delta x_k^{(1)} > 0
\label{eq:deltaq}
\end{equation}

We must have $\partial R_k / \partial B_k > 0$, since the revenue increases
with $B_k$ when $x_k$ is fixed, hence optimizing $x_k$ can only increase
the revenue further. 
%We can compute
%\begin{equation}
%\frac{\partial R_{-k}}{\partial B_{k}} = x_{-k} P' (x_1+x_2) \frac{\partial x_k}{\partial B_k}
%\end{equation}
%which is negative.
The incremental revenue for agent $-k$ is given by
\begin{align}
\nonumber
\delta R_{-k} & = x_{-k} P' (x_1+x_2) (\delta x_{k,eq} + \delta x_{-k,eq})
+ [P(x_1 + x_2) - \ell_{-k} (x_{-k}/B_{-k})] \delta x_{-k,eq}  \\
& ~~~
-\frac{x_{-k}}{B_{-k}} \ell_{-k}' (x_{-k}/B_{-k})  \delta x_{-k,eq}
\end{align}
where the first term ($< 0$) is due to the reduction in total price,
the second term ($< 0$) is due to the reduced quantity at the announced price,
and the last term ($> 0$) is due to reduction in latency.
Combining with the best response condition \eqref{eq:br}, this simplifies to
\begin{equation}
\delta R_{-k} = x_{-k} P' (x_1 + x_2) \cdot \delta x_{k,eq} < 0.
\end{equation}

\subsection{Proof of Proposition \ref{marg:prop2} }
To prove Proposition \ref{marg:prop2}, we note that
the increase in consumer surplus follows directly
from \eqref{eq:deltaq}. The increase in total welfare can be shown
by adding the incremental areas when $x_k$ and $x_{-k}$ are
incremented by $\delta x_{k,eq}$ and $\delta x_{-k,eq}$.
(can add this later...) 

To determine which agent should receive the bandwidth to generate 
the most consumer surplus, we maximize the incremental quantity in \eqref{eq:deltaq}.
Equivalently, we wish to determine
\begin{equation}
\arg \max_k (1+ x_{-k,k}) \frac{\partial x_k}{\partial B_k}
\end{equation}
Substituting for the derivatives, this becomes
\begin{equation}
\arg \max \left( P' (x_1+p_2) - \frac2B_{-k} \ell_{-k}' (\bar{x}_{-k}) 
- \frac1{B_{-k}^2} \ell_{-k}'' (\bar{x}_{-k} ) \right)
\left( \frac{2 x_k}{B_k^2} \ell_2 (\bar{x}_k ) + 
\frac{x_k^2}{B_k^3} \ell_k'' (\bar{x}_k ) \right)
\end{equation}
where $\bar{x}_k = x_k/B_k$. For linear latencies $\ell_k (x) = c_k x$
this reduces to finding
\begin{equation}
\arg \max - \frac{c_k x_k}{B_k^2} \left( P' (x_1 + x_2 ) ,
- \frac{2 c_{-k}}{B_{-k}} \right) 
\end{equation}
which this reduces to \eqref{eq:marg1},
and further constraining $P(x) = 1 - ax$ gives the condition \eqref{eq:marg2}.

%%%%%%%%%%%%%%%%%%%%%
%%%%%%%%%%%%%%%%%%%%%
%%%%%%%%%%%%%%%%%%%%%

\Xomit{\color{blue}
when there are $n$ SPs

\begin{align*}
x (n+1+\tfrac{2}{K}) + w (n+1+\tfrac{(1-\alpha)(1+\beta)}{K}) & =1 \\
x (n+1+\tfrac{(1-\alpha)(1+\beta)}{K}) + w (n+1+\tfrac{(n+1)\alpha}{W}+\tfrac{2 (1-\alpha)\beta}{K}) & =1 
\end{align*}

from the two equation we will get

\begin{equation}\label{eq:x/w}
\frac{x}{K}=\frac{w}{W}\frac{n+1}{2}
\end{equation}

There is an interpretation here!!! it is funny that it does not depend on $\alpha$\\

\begin{itemize}
\item Variables  of interests:  
$$nx; nw$$
\item we analize the  behavior of  equilibrium accoring to  the following parameters 
$$W; \frac{W}{nK}:=\Delta$$
\end{itemize}

We will solve the equation above as $n$ goes to infinity.

from \eqref{eq:x/w} we have
$$
nw=2\Delta nx \text{ explain the  interpretation.....}
$$

$$
nx=\frac{1}{1+2\Delta+(1+2\Delta(1-\alpha))\frac{2\Delta}{W}} \text{ explain the  interpretation.....}
$$
$$
nw=\frac{2\Delta}{1+2\Delta+(1+2\Delta(1-\alpha))\frac{2\Delta}{W}} \text{ explain the  interpretation.....}
$$
total traffic is 
$$
\boxed{
nx+nw=\frac{1+2\Delta}{1+2\Delta+(1+2\Delta -2\Delta\alpha)\frac{2\Delta}{W}}
}
  \text{ explain the  interpretation.....}
$$

now we compare this with  divive $W$ equally, by similar analysis;
we obtain total traffic is

$$
\boxed{
nx=\frac{1+\Delta}{1+\Delta+(1+ \Delta-\Delta\alpha)\frac{2\Delta}{W}}
}
$$

}

\end{document}